\documentclass[12pt,a4paper, amsmath,amssymb]{article}
\usepackage[margin=0.8in]{geometry}
\usepackage{amssymb}
\usepackage{amsmath}
\usepackage{amsfonts}
\usepackage[utf8]{inputenc}
\usepackage{graphicx}
\usepackage{float}
\usepackage{xcolor}
\usepackage{caption}
\definecolor{bluemunsell}{rgb}{0.0, 0.5, 0.69}
\usepackage{float}
\usepackage{adjustbox}
\usepackage{relsize}
\usepackage[colorlinks,linkcolor = bluemunsell,
urlcolor  = bluemunsell,
citecolor = bluemunsell,
anchorcolor = bluemunsell]{hyperref}\usepackage{hyperref}
\usepackage[normalem]{ulem}

\allowdisplaybreaks
\usepackage{xcolor}
\usepackage{makecell}
\usepackage{multirow}
\usepackage{tikz}

\usepackage{cite}
\usepackage{tikz}
\usepackage{ctable}
\usepackage{feynmp-auto}
\definecolor{green}{rgb}{0.0, 0.56, 0.0}

\providecommand{\openone}{\leavevmode\hbox{\small1\kern-3.8pt\normalsize1}}

\newcommand{\tb}{$(TB)$}
\newcommand{\tby}{$(TBY)$}
\def\half{\tfrac{1}{2}}
\definecolor{green}{rgb}{0.0, 0.56, 0.0}

\usepackage{etoolbox}
\makeatletter

\providecommand{\openone}{\leavevmode\hbox{\small1\kern-3.8pt\normalsize1}}

\newcommand{\xtb}{$(XTB)$}
\def\half{\tfrac{1}{2}}

\makeatother
	
\begin{document}
		
		\begin{center}
			{\Large \textbf{Anatomy of Vector-Like Bottom-Quark Models in the \\\vspace{0.2cm}
				Alignment Limit of the 2-Higgs Doublet Model Type-II\\\vspace{0.2cm} 
			}}
			\thispagestyle{empty}
			\def\thefootnote{\fnsymbol{footnote}}
			\vspace{1cm}
			
			{\sc
				A. Arhrib$^{1,2}$\footnote{\href{mailto:aarhrib@gmail.com}{aarhrib@gmail.com}},
				R. Benbrik$^3$\footnote{\href{mailto:r.benbrik@uca.ac.ma}{r.benbrik@uca.ac.ma}},
				M. Boukidi$^3$\footnote{\href{mailto:mohammed.boukidi@ced.uca.ma}{mohammed.boukidi@ced.uca.ma}},
				S. Moretti$^{4,5}$\footnote{\href{mailto:stefano.moretti@physics.uu.se}{stefano.moretti@physics.uu.se}; \href{mailto:s.moretti@soton.ac.uk}{s.moretti@soton.ac.uk}}\\}
	
			\vspace{1cm}
			{\sl\small
				$^1$Abdelmalek Essaadi University, Faculty of Sciences and Techniques, Tangier, Morocco\\
						\vspace{0.1cm}
						
					$^2$	Department of Physics and CTC, National Tsing Hua University, Hsinchu 300, Taiwan\\		
				$^3$Polydisciplinary Faculty, Laboratory of Fundamental and Applied Physics, Cadi Ayyad University, Sidi Bouzid, B.P. 4162, Safi, Morocco\\
				\vspace{0.1cm}
				$^4$Department of Physics \& Astronomy, Uppsala University, Box 516, SE-751 20 Uppsala, Sweden\\
				
				\vspace{0.1cm}
				$^5$School of Physics \& Astronomy, University of Southampton, Southampton, SO17 1BJ, United Kingdom
			}
		\end{center}
		\vspace*{0.1cm}
		\begin{abstract}
			
			Expanding upon our ongoing investigation of Vector-Like Quark (VLQ) phenomenology within a 2-Higgs Doublet Model (2HDM) framework, in this paper, we complement a previous one dedicated to Vector-Like Top-quarks (VLTs) by studying Vector-Like Bottom-quarks (VLBs), specifically focusing on their behavior in the alignment limit of a Type-II Yukawa structure. We examine the potential for detecting VLBs at the Large Hadron Collider (LHC) and analyze their decay signatures, encompassing both Standard Model (SM) processes and exotic decays. The objective is to differentiate among singlet, doublet, and triplet configurations of VLBs by identifying distinct decay patterns, thereby providing insights into the structure of Beyond the SM (BSM) physics.  
			
		\end{abstract}
		
		\def\thefootnote{\arabic{footnote}}
		\setcounter{page}{0}
		\setcounter{footnote}{0}

		\newpage
		
		\section{Introduction} 

Building on the remarquable  discovery of the Higgs boson during the LHC Run 1 at CERN and the  subsequent detailed analyses of its properties by the ATLAS and CMS collaborations \cite{ATLAS:2012yve, CMS:2012qbp}, our focus here is on exploring Vector-Like Quarks in depth, particularly focusing on the Bottom-type VLQs (VLBs), in this continuation of our previous research \cite{Arhrib:2024tzm}.  The SM predictions have been largely validated by the identification of the Higgs boson,
despite this, anomalies observed in various production and decay channels suggest the potential for physics beyond the SM (BSM). 
It is widely acknowledged that a lot of BSM models necessitate the use of new fermions, or new scalar bosons, or both. This drives us to examine the role of VLBs in greater depth.

In the context of the 2-Higgs Doublet Model (2HDM) \cite{Branco:2011iw,Draper:2020tyq}, which introduces additional Higgs bosons ($H$, $A$, and $H^\pm$) in addition to the discovered SM-like one ($h$), our study aims to uncover the implications of including VLBs with unique Electro-Weak (EW) quantum numbers and non-Yukawa type couplings.
The purpose of this research is to expand the analysis that has been done recently on Top-type VLQs (VLTs) \cite{Arhrib:2024tzm}. By studying the exotic decay channels of VLBs into new Higgs states $H$, $A$, and $H^\pm$, a subject that has not been extensively explored at the LHC.

Our motivation comes from the possibility of VLBs decaying into one of the  additional Higgs bosons $H$, $A$, and $H^\pm$ together with a quark like 
$B\to \{ tH^\pm, bH, bA\}$, this could be employed to avoid detection by current experimental search strategies at the LHC, 
which are primarily focused on SM decay modes such as $B\to \{ tW, Zb, bh\}$, and to assess the model alignment with EW Precision Observables (EWPOs). 
Through this work, we aim to contribute to a deeper understanding of VLBs within the BSM framework, 
emphasizing the importance of their unique decay channels in revealing novel aspects of particle physics.

This paper closely follows  a similar one of ours on VLTs \cite{Arhrib:2024tzm}  and we made the decision to separate them to align with the methodology 
used by experimentalists at the LHC. The ATLAS and CMS collaborations search for VLQs by their signature, thus focusing on different ones for $T$-type or $B$-type VLQs.
This separation allows for targeted data collections and analysis strategies tailored to the unique characteristics and interactions of each VLQ type. By producing separate publications for VLTs and VLBs,  we ensure that our research provides specific inputs (e.g., in the form of theoretical scenarios to probe using different parameter configurations for VLTs and VLBs) that are directly applicable to ongoing  experimental work on VLQs.

In this paper, we expand the literature series on VLQs in the context of  2HDM type II  \cite{Benbrik:2015fyz,Arhrib:2016rlj,Aguilar-Saavedra:2013qpa,Badziak:2015zez,Angelescu:2015uiz,Aguilar-Saavedra:2009xmz,DeSimone:2012fs,Kanemura:2015mxa,Lavoura:1992np,Chen:2017hak,Carvalho:2018jkq,Moretti:2016gkr,Prager:2017hnt,Prager:2017owg,Moretti:2017qby,Deandrea:2017rqp,Aguilar-Saavedra:2017giu,Alves:2023ufm,Dermisek:2019vkc,Dermisek:2020gbr,Dermisek:2021zjd,Benbrik:2022kpo,Vignaroli:2012sf,Vignaroli:2015ama,Vignaroli:2012si}.
We first elaborate on the theoretical foundations of the 2HDM and VLBs, subsequently, a comprehensive examination of the decay patterns of VLBs and their potential effects on LHC detection and characterization. Using knowledge from our previous work on VLTs, particularly the 2HDM Type-II \cite{Arhrib:2024tzm}, 
our aim is to offer a comprehensive perspective on the capabilities of VLBs to illuminate new physics phenomena.

The paper's structured  as follows. In the next section, we discuss the 2HDM augmented by VLQs and present our parameterization for different VLQ representations: singlet, doublet and triplet. In the following section, we present and discuss our results. 
Then we summarize our work and draw our conclusions.

\section{Model description}

\subsection{Formalism}
Our paper builds on the framework that was introduced in Ref. \cite{Arhrib:2024tzm}, which examines a 2HDM Type-II with VLQ representations. 
Wherein the phenomenology of this BSM setup was accessed via VLT decay dynamics, while here we will establish the one of the VLB sector.

The physical spectrum of the 2HDM contains the following set of Higgs bosons: two CP-even states, with $h$ being the lightest and $H$ the heaviest, a CP-odd state, $A$, and a pair of charged Higgs bosons, $H^\pm$. It is well known that flavor-changing neutral currents (FCNCs) at the tree-level are highly constrained by experimentation. Therefore, 
to avoid having FCNC at the tree level,  we impose a $\mathbb{Z}_2$ symmetry, such that $\Phi_1 \to \Phi_1$ and $\Phi_2 \to - \Phi_2$, on the Higgs fields. The resulting Higgs potential which conserve $\mathbb{Z}_2$ symmetry  can be written as 
\begin{eqnarray} \label{pot}
\mathcal{V} &=& m^2_{11}\Phi_1^\dagger\Phi_1+m^2_{22}\Phi_2^\dagger\Phi_2
-\left(m^2_{12}\Phi_1^\dagger\Phi_2+{\rm h.c.}\right)
\nonumber \\
&&+\half\lambda_1\left(\Phi_1^\dagger\Phi_1\right)^2
+\half\lambda_2\left(\Phi_2^\dagger\Phi_2\right)^2  +\lambda_3\Phi_1^\dagger\Phi_1\Phi_2^\dagger\Phi_2
\nonumber \\
&&+\lambda_4\Phi_1^\dagger\Phi_2\Phi_2^\dagger\Phi_1
+\left[\half\lambda_5\left(\Phi_1^\dagger\Phi_2\right)^2+{\rm h.c.}\right].
\end{eqnarray}
Note that   $\mathbb{Z}_2$ is softly broken by  dimension-2 terms $\propto m^2_{12}$ only.
By choosing real Vacuum Expectation Values (VEVs) for the two Higgs doublet fields, $v_1$ and $v_2$, and demanding
$m_{12}^2$ and $\lambda_5$ to be real, the potential is  CP-conserving.
The independent parameters of the scalar sector of the 2HDM include the masses $m_h$, $m_H$, $m_A$, and $m_{H^\pm}$, the soft $\mathbb{Z}2$-breaking parameter $m{12}$, the VEV ratio $\tan \beta = v_2/v_1$, and the mixing term $\sin(\beta-\alpha)$, where the angle $\alpha$ diagonalizes the CP-even mass matrix.
When we impose that no significant tree-level FCNCs are present in the theory using the (softly broken) $\mathbb{Z}_2$ symmetry, 
this results in four possible Yukawa configurations: Type-I (where only $\Phi_2$ interacts with all fermions), Type-II (where $\Phi_2$ couples to up-type quarks and $\Phi_1$ to down-type quarks and charged leptons), Type-Y (or Flipped) where $\Phi_2$ couples to charged leptons and up-type quarks, and $\Phi_1$ to down-type quarks, and Type-X (or Lepton Specific) where $\Phi_2$ couples to quarks, and $\Phi_1$ to charged leptons\footnote{Our focus in this paper is on Type-II.}.
The gauge invariant structures that have multiplets with definite ${SU}(3)_C \times {SU}(2)_L \times {U}(1)_Y$ quantum numbers appear in the interactions of new VLQs with the SM states via renormalizable couplings. The set of VLQ representations is indicated by:

\begin{align}
& B_{L,R}^0 \, && \text{(singlets)} \,, \notag \\
& (B^0\,Y)_{L,R} \,, \quad (T^0\,B^0)_{L,R} \, && \text{(doublets)} \,, \notag \\
& (X\,T^0\,B^0)_{L,R} \,, \quad (T^0\,B^0\,Y)_{L,R}  && \text{(triplets)} \,.
\end{align}
In this section, we differentiate the mass eigenstates from the weak eigenstates using a zero superscript. 
The Electro-Magnetic (EM)  charges of the new VLQs are $ Q_T = 2/3 $, $ Q_B = -1/3 $, $ Q_X = 5/3 $ and $ Q_Y = -4/3 $. 
Take note that, for the top and bottom quarks in the SM, $T$ and $B$ have the same EM charge, respectively.

The physical up-type quark mass eigenstates may, in general, contain non-zero $Q_{L,R}^0$ (with $Q$ being the VLQ field) components, when new fields $T_{L,R}^0$ of charge $2/3$ and non-standard isospin assignments are added to the SM. This situation leads to a deviation in their couplings to the $Z$ boson. Atomic parity violation experiments and the measurement of $R_c$ at LEP impose constraints  on these deviations for the up and charm quarks which are far stronger than for the top quark.
In the Higgs basis, the Yukawa Lagrangian contains the following terms:
\begin{equation}
-\mathcal{L} \,\, \supset  \,\, y^u \bar{Q}^0_L \tilde{H}_2 u^0_R +  y^d \bar{Q}^0_L H_1 d^0_R + M^0_u \bar{u}^0_L u^0_R  + M^0_d \bar{d}^0_L d^0_R + \rm {H.c.}
\label{lag-yuk}
\end{equation}
Here, $u_R$  runs over $(u_R, c_R, t_R, T_R)$ while $d_R$  runs over \\ $(d_R, s_R, b_R, B_R)$. 

We now turn to the mixing of the new fermionic partners to the third generation, $y_u$ and $y_d$, which are $3\times 4$ Yukawa matrices. 
It is very reasonable to assume that only the top quark $t$  ``mixes'' with $T$.
In this case,  the relationship between the charge $2/3$ weak and mass eigenstates is through  the $2 \times 2$ unitary matrices $U_{L,R}^u$ as:
\begin{eqnarray}
\left(\! \begin{array}{c} t_{L,R} \\ T_{L,R} \end{array} \!\right) &=&
U_{L,R}^u \left(\! \begin{array}{c} t^0_{L,R} \\ T^0_{L,R} \end{array} \!\right)
\nonumber \\
&=& \left(\! \begin{array}{cc} \cos \theta_{L,R}^u & -\sin \theta_{L,R}^u e^{i \phi_u} \\ \sin \theta_{L,R}^u e^{-i \phi_u} & \cos \theta_{L,R}^u \end{array}
\!\right)
\left(\! \begin{array}{c} t^0_{L,R} \\ T^0_{L,R} \end{array} \!\right) \,.
\label{ec:mixu}
\end{eqnarray}
In contrast to the up-type quark sector, the addition of new fields $B_{L,R}^0$ of charge $-1/3$ in the down-type quark sector results in four mass eigenstates, $d,s,b,B$.
Measurements of $R_b$ at LEP set severe constraints on the $b$ mixing with the new fields that are stronger than for mixing with the lighter quarks $d$ and $s$. 
In this case, then,  $2 \times 2$ unitary matrices $U_{L,R}^d$ define the dominant $b-B$ mixing as 
\begin{eqnarray}
\left(\! \begin{array}{c} b_{L,R} \\ B_{L,R} \end{array} \!\right)
&=& U_{L,R}^d \left(\! \begin{array}{c} b^0_{L,R} \\ B^0_{L,R} \end{array} \!\right)
\nonumber \\
&=& \left(\! \begin{array}{cc} \cos \theta_{L,R}^d & -\sin \theta_{L,R}^d e^{i \phi_d} \\ \sin \theta_{L,R}^d e^{-i \phi_d} & \cos \theta_{L,R}^d \end{array}
\!\right)
\left(\! \begin{array}{c} b^0_{L,R} \\ B^0_{L,R} \end{array} \!\right) \,.
\label{ec:mixd}
\end{eqnarray}
To ease the notation, we have dropped the superscripts $u$($d$) whenever the mixing occurs only in the up(down)-type quark sector. 
Additionally, we sometime use the shorthand notations $s_{L,R}^{u,d} \equiv \sin \theta_{L,R}^{u,d}$, $c_{L,R}^{u,d} \equiv \cos \theta_{L,R}^{u,d}$, etc.\\
The above Lagrangian eq.\ref{lag-yuk}  contains all the  relevant phenomenological  information, as follows:  
\begin{itemize}
	\item[(i)] the modifications of the SM couplings that might show indirect effects of new quarks can be found in the terms that do not contain VLQ fields,
	\item[(ii)] the terms relevant for LHC phenomenology (i.e., VLQ production and decay) are those involving a heavy and a light quark,
	\item[(iii)] terms with two VLQs are relevant for their contribution to oblique corrections (i.e., to the $S,T$ and $U$ parameters of the EWPOs).
\end{itemize}
More details on this Lagrangian  formalism given in eq.\ref{lag-yuk} are shown in the Appendix.

In the weak eigenstate basis, the diagonalization of the mass matrices makes the Lagrangian of the third generation and heavy quark mass terms such as
\begin{eqnarray}
\mathcal{L}_\text{mass} & = & - \left(\! \begin{array}{cc} \bar t_L^0 & \bar T_L^0 \end{array} \!\right)
\left(\! \begin{array}{cc} y_{33}^u \frac{v}{\sqrt 2} & y_{34}^u \frac{v}{\sqrt 2} \\ y_{43}^u \frac{v}{\sqrt 2} & M^0 \end{array} \!\right)
\left(\! \begin{array}{c} t^0_R \\ T^0_R \end{array}
\!\right) \notag \\
& & - \left(\! \begin{array}{cc} \bar b_L^0 & \bar B_L^0 \end{array} \!\right)
\left(\! \begin{array}{cc} y_{33}^d \frac{v}{\sqrt 2} & y_{34}^d \frac{v}{\sqrt 2} \\ y_{43}^d \frac{v}{\sqrt 2} & M^0 \end{array} \!\right)
\left(\! \begin{array}{c} b^0_R \\ B^0_R \end{array}
\!\right) +\text{H.c.},
\label{ec:Lmass}
\end{eqnarray}
Where  $y_{ij}^q$, $q=u,d$, Yukawa couplings and  $v=246$ GeV the Higgs VEV in the SM and  $M^0$ a bare mass term. Note that this bare mass term is not related to the Higgs mechanism: it is gauge invariant and can appear as such in the Lagrangian, or it can be generated by a Yukawa coupling to a scalar multiplet that acquires a VEV $v' \gg v$.

The mixing matrices can be obtained by utilizing standard diagonalization techniques
\begin{equation}
U_L^q \, \mathcal{M}^q \, (U_R^q)^\dagger = \mathcal{M}^q_\text{diag} \,,
\label{ec:diag}
\end{equation}
with $\mathcal{M}^q$ the two mass matrices in Eq.~(\ref{ec:Lmass}) and $\mathcal{M}^q_\text{diag}$ the diagonals ones. 
To check the consistency of our calculation, the corresponding $2 \times 2$ mass matrix reduces to the SM quark mass term if either the $T$ or $B$ quarks are absent.

It should be noted that when multiplets contain both $T$ and $B$ quarks, the bare mass value remains identical for both the up- and down-type quark sectors.
For singlets and triplets one has $y_{43}^q = 0$ whereas for doublets $y_{34}^q=0$. Moreover, for the \xtb\ triplet, one has $y_{34}^d = \sqrt 2 y_{34}^u$ and for the \tby\ triplet one has $y_{34}^u = \sqrt 2 y_{34}^d$\footnote{We write the triplets in the spherical basis, hence, the $\sqrt 2$ factors stem from the relation between the Cartesian and spherical coordinates of an irreducible tensor operator of rank 1 (vector).}.

The mixing angles in the left- and right-handed sectors are not independent parameters. From the mass matrix bi-unitary diagonalization in Eq.~(\ref{ec:diag}) one finds:
\begin{eqnarray}
\tan 2 \theta_L^q & = & \frac{\sqrt{2} |y_{34}^q| v M^0}{(M^0)^2-|y_{33}^q|^2 v^2/2 - |y_{34}^q|^2 v^2/2} \quad \text{(singlets, triplets)} \,, \notag \\
\tan 2 \theta_R^q & = & \ \frac{\sqrt{2}  |y_{43}^q| v M^0}{(M^0)^2-|y_{33}^q|^2 v^2/2 - |y_{43}^q|^2 v^2/2} \quad \text{(doublets)} \,,
\label{ec:angle1}
\end{eqnarray}
with the relations
\begin{eqnarray}
\tan \theta_R^q & = & \frac{m_q}{m_Q} \tan \theta_L^q \quad \text{(singlets, triplets)} \,, \notag \\ \\
\tan \theta_L^q & = & \frac{m_q}{m_Q} \tan \theta_R^q \quad \text{(doublets)} \,,\notag
\label{ec:rel-angle1}
\end{eqnarray}
One mixing angle is always the predominant one, particularly in the down-type quark sector.

with $(q,m_q,m_Q) = (u,m_t,m_T)$ and $(d,m_b,m_B)$, so one mixing angle is always the predominant one, particularly in the down-type quark sector.
In addition, for the triplets, the relations between the off-diagonal Yukawa couplings lead to relations between the mixing angles in the up- and down-type quark  sectors,
\begin{eqnarray}
\sin 2\theta_L^d & = & \sqrt{2} \, \frac{m_T^2-m_t^2}{m_B^2-m_b^2} \sin 2 \theta_L^u \quad \quad (X\,T\,B) \,, \notag \\
\sin 2\theta_L^d & = & \frac{1}{\sqrt{2}} \frac{m_T^2-m_t^2}{m_B^2-m_b^2} \sin 2 \theta_L^u \quad \quad (T\,B\,Y) \,.
\label{TBY-mix}
\end{eqnarray}

The masses of the heavy VLQs deviate from $M^0$ due to the non-zero mixing with the SM quarks, plus, for doublets and triplets, the masses of the different components of the multiplet are related. Altogether, these relations show that all multiplets except the \tb\ doublet can be parametrized by a mixing angle, a heavy quark mass and a CP-violating phase that enters some couplings, with the latter being ignored for the observables considered in this paper. 
In the case of the \tb\ doublet, there are two independent mixing angles and two CP-violating phases for the up- and down-type quark sectors, with - again - the latter set to zero in our analysis. Hereafter, we refer to such a construct as the 2HDM+VLQ scenario,  distinguishing between the singlet, doublet  and triplet cases. In the present paper, though, given the emphasis on VLBs, as opposed to VLTs, we will treat the $(B)$ singlet,   $(BY)$ and $(TB)$ doublets as well as $(XTB)$ and $(TBY)$ triplets whereas we will not deal with the $(T)$ singlet and $(XT)$ doublet representations, as their study was tackled in Ref.~\cite{Arhrib:2024tzm}. Finally, as discussed in the abstract, we decided to work in the so-called ``alignment limit'' of the 2HDM, wherein we fix $m_h=125$ GeV (so that the lightest neutral Higgs state of the 2HDM is the discovered one) and we have further taken $m^2_{12}=m_A^2\frac{\tan^2\beta}{1+\tan^2\beta}$ for the soft $\mathbb{Z}_2$-breaking parameter.

\subsection{Model Implementation and Validation}

In this subsection, we outline the implementation of our BSM scenario. We utilized \texttt{2HDMC-1.8.0} \cite{Eriksson:2009ws} as the foundational platform for our 2HDM+VLQ setup\footnote{A public release of it is in progress:  herein, the analytical expressions for the Feynman rules of the interaction vertices of our 2HDM+VLQ model have been implemented as a new class while several new tree-level VLQ decays have explicitly been coded alongside those of Higgs bosons into VLQs themselves.}. Initially, the Lagrangian components were implemented into \texttt{FeynRules-2.3} \cite{Degrande:2011ua} to generate proper spectrum of masses and couplings. Using this tool, we produced \texttt{FeynArts-3.11} \cite{Hahn:2000kx,Kublbeck:1990xc} and \texttt{FormCalc-9.10} \cite{Hahn:2001rv,Hahn:1998yk} model files, as well as Universal FeynRules Output (UFO) interfaces for \texttt{MadGraph-3.4.2} \cite{Alwall:2014hca}. For consistency checks, we verified the cancellation of ultraviolet (UV) divergences and the renormalization scale independence of all relevant loop-level processes.

\subsection{Constraints}
\label{sec-A}
In this section, we list the constraints that we have used to check the validity of our results.
From the theoretical side, we have the following requirements:
\begin{itemize}
	\item \textbf{Unitarity} constraints require the $S$-wave component of the various
	(pseudo)scalar-(pseudo)scalar, (pseudo)scalar-gauge boson, and gauge-gauge bosons scatterings to be unitary
	at high energy ~\cite{Kanemura:1993hm}.
	\item \textbf{Perturbativity} constraints impose the following condition on the quartic couplings of the scalar potential: $|\lambda_i|<8\pi$ ($i=1, ...5$)~\cite{Branco:2011iw}.    
	\item \textbf{Vacuum stability} constraints require the potential to be bounded from below and positive in any arbitrary
	direction in the field space, as a consequence, the $\lambda_i$ parameters should satisfy the conditions as~\cite{Barroso:2013awa,Deshpande:1977rw}:
	\begin{align}
	\lambda_1 > 0,\quad\lambda_2>0, \quad\lambda_3>-\sqrt{\lambda_1\lambda_2} ,\nonumber\\ \lambda_3+\lambda_4-|\lambda_5|>-\sqrt{\lambda_1\lambda_2}.\hspace{0.5cm}
	\end{align} 
	
	\item \textbf{Constraints from EWPOs}, implemented through the oblique parameters, $S$ and $T$ ~\cite{Grimus:2007if},  require that, for a parameter point of our
	model to be allowed, the corresponding $\chi^2(S^{\mathrm{2HDM\text{-}II}}+S^{\mathrm{VLQ}},~T^{\mathrm{2HDM\text{-}II}}+T^{\mathrm{VLQ}})$ is within 95\% Confidence Level (CL) in matching the global fit results \cite{Molewski:2021ogs}:
	\begin{align}
	&S= 0.05 \pm 0.08,\quad T = 0.09 \pm 0.07,\nonumber \\&  \rho_{S,T} = 0.92 \pm 0.11.\hspace{0.5cm}(\text{For~~}U=0) 
	\end{align}
	{A comprehensive discussion on EWPOs contributions in VLQs can be found in \cite{Arhrib:2024tzm, Abouabid:2023mbu}.} Note that unitarity, perturbativity, vacuum stability, as well as $S$ and $T$ constraints, are enforced through the public code \texttt{2HDMC-1.8.0} \cite{Eriksson:2009ws}.
\end{itemize}
From the experimental side, we evaluated the following:
\begin{itemize}
	\item \textbf{Constraints from the SM-like Higgs-boson properties}  are taken into account by using \texttt{HiggsSignal-3} \cite{Bechtle:2020pkv,Bechtle:2020uwn} via \texttt{HiggsTools} \cite{Bahl:2022igd}. We require that the relevant quantities (signal strengths, etc.) satisfy $\Delta\chi^2=\chi^2-\chi^2_{\mathrm{min}}$ for these measurements at 95\% CL ($\Delta\chi^2\le6.18$), involving 159 observables.
	\item\textbf{Constraints from direct searches at colliders}, i.e., LEP, Tevatron, and LHC, are taken at the 95\% CL and are tested using \texttt{HiggsBouns-6}\cite{Bechtle:2008jh,Bechtle:2011sb,Bechtle:2013wla,Bechtle:2015pma} via \texttt{HiggsTools}. Including the most recent searches for neutral and charged scalars.
	
	\item {\bf Constraints from $b\to s\gamma$}: To align with the $b\to s\gamma$ limit, the mass of the charged Higgs boson is set to exceed 600 GeV. This is based on the analysis in Ref \cite{Benbrik:2022kpo}, which indicates that incorporating VLQs into the Type-II 2HDM could relax this limit through larger mixing angles. However, EWPOs constrain these angles to smaller values, consequently maintaining the charged Higgs mass close to the typical 580 GeV in the 2HDM Type-II.
	
\end{itemize}
\subsection{Direct search constraints}
In this section, we explore the implications of experimental constraints on the properties of additional Higgs states and VLQs within our 2HDM+VLQ framework. The oblique parameters $S$ and $T$ play a crucial role by imposing stringent limits on the VLQ mixing angles, as elaborated in our previous work \cite{Benbrik:2022kpo} and in the previous subsection. This limitation leads to small mixing angles, allowing VLQ masses to start from as low as 1000 GeV without conflicting with existing LHC exclusions. This pattern holds true across singlet, doublet and triplet VLQ configurations.

We then provide an illustrative example for  both the singlet 2HDM+$B$ and the doublet 2HDM+$BY$ scenarios, depicting the consistency of our findings against the LHC limits. The blue points representing our results that satisfy all the discussed constraints in the $(m_B, \kappa)$ plane are overlaid on the 95\% Confidence Level (CL) observed (solid) and expected (dashed) contours from ATLAS\footnote{It is important to note that, despite the apparent consistency of our data points with the LHC limits, we have deliberately incorporated these limits into our code. This approach allows us to thoroughly evaluate each point against the 95\% CL exclusion criteria. By doing so, we ensure that only points fully compliant with established thresholds are retained, thereby reinforcing the rigor and reliability of our analysis.} \cite{ATLAS:2023ixh}. Recalling our earlier discussion on the constraint imposed by the oblique parameters $S$ and $T$ in constraining the VLQ mixing angles to small values, it becomes evident from the figure that our results consistently reside below the observed 95\%  CL exclusion limit.
\begin{figure*}[t!]
	\centering
	\includegraphics[width=0.85\textwidth,height=0.40\textwidth]{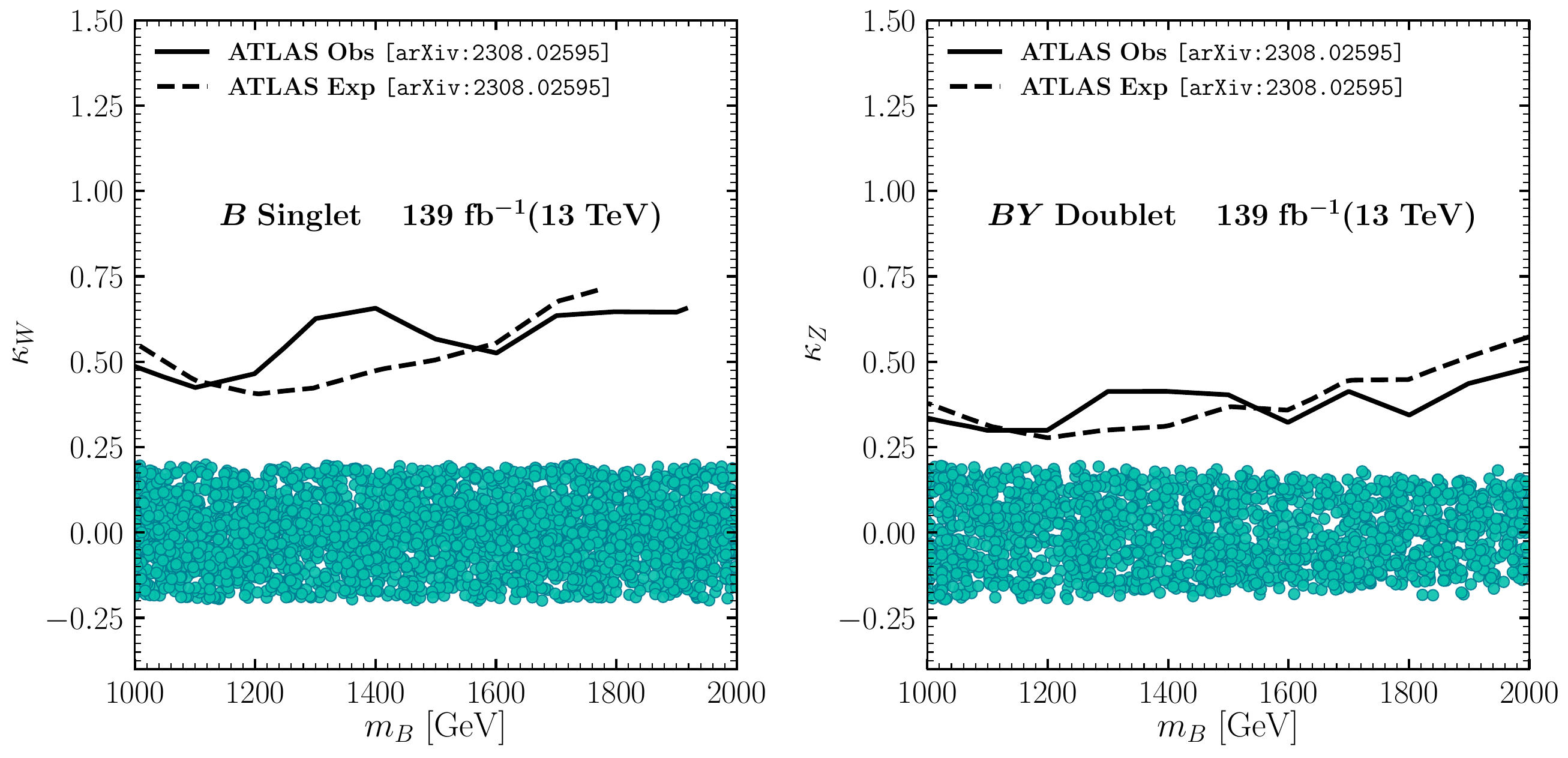}
	\caption{Allowed points following the discussed theoretical and experimental constraints, in the ($m_B, \kappa$) plane for for the 2HDM+$B$ singlet scenario (left) and 2HDM+$BY$ doublet scenario (right), superimposed onto the ATLAS \cite{ATLAS:2023ixh} 95\% C.L observed (solid line) and expected (dashed) upper limits on the coupling $\kappa$.}
	\label{fig_1}
\end{figure*}

\section{VLQ Effects on Higgs Production and Decay}

VLQs have an impact on Higgs production through gluon-gluon fusion and its decay into two photons. In our analysis, these contributions are minor. This is due to two primary reasons: VLQs tend to decouple as their masses increase, and the mixing angles $({s^{d}_{L,R}})^2$ present in the $hB\overline B$ coupling (see Table XIV in the appendix) are constrained by EWPOs, resulting in small values further diminished by their squared terms. The main, though small, contribution stems from changes in the $hb\bar b$ coupling.

The contributions of all quarks with the same charge to the amplitudes for processes like $gg\to H$ (or $H \to gg$) and $H \to \gamma \gamma$ are determined by the expression:
\begin{equation}
F_q=\sum_i Y_{ii} A_{1/2}\left(\frac{M_H^2}{4 m_i^2} \right),
\end{equation}
where the sum includes $t,T$ for $q=u$ and $b,B$ for $q=d$. The Higgs couplings $Y_{ii}$ are defined in Table XIV of the appendix, and the function $A_{1/2}$ is detailed in~\cite{Djouadi:2005gi}.

Our findings indicate that, across various scenarios, the branching ratio $\mathcal{BR}(h\to gg)$ is reduced by up to 10\%, primarily due to changes in the $hb\bar b$ coupling. Similarly, the branching ratio $\mathcal{BR}(h\to \gamma\gamma)$ decreases by up to 3\%, also due to modifications in the $hb\bar b$ coupling. The new terms $hB\overline B$ have negligible contributions, as previously discussed.

\section{Results and discussions}
In discussing our results, we deal with these in turn for the different VLQ representations.

\subsection{2HDM with $(B)$ singlet}
\begin{table}[t!]
	\centering
	{\setlength{\tabcolsep}{1.5cm}
		\begin{tabular}{cc}
			\toprule\toprule
			Parameters  & Scanned ranges \\
			\toprule
			$m_h$   & $125$ \\
			$m_A$  & [$200$, $1000$] \\
			$m_H$  & [$200$, $1000$] \\
			$m_{H^\pm}$  & [$600$, $1000$] \\
			$\tan\beta$ & [$1$, $20$] \\
			$m_{B}$   & [$1000$, $2000$] \\	
			$\sin\theta_L^{u,d}$  & [$-0.8$, $0.8$] \\
			$\sin\theta_R^{u,d}$  & [$-0.8$, $0.8$] \\
			\toprule\toprule
			
	\end{tabular}}
	\caption{2HDM and VLQs parameters for all scenarios with their scanned ranges. Masses are in GeV. The phases $\phi_u$ and $\phi_d$ are set to zero.}
	\label{table1}
\end{table}
\begin{figure}[t!]
	\centering
	\includegraphics[width=0.85\textwidth,height=0.4\textwidth]{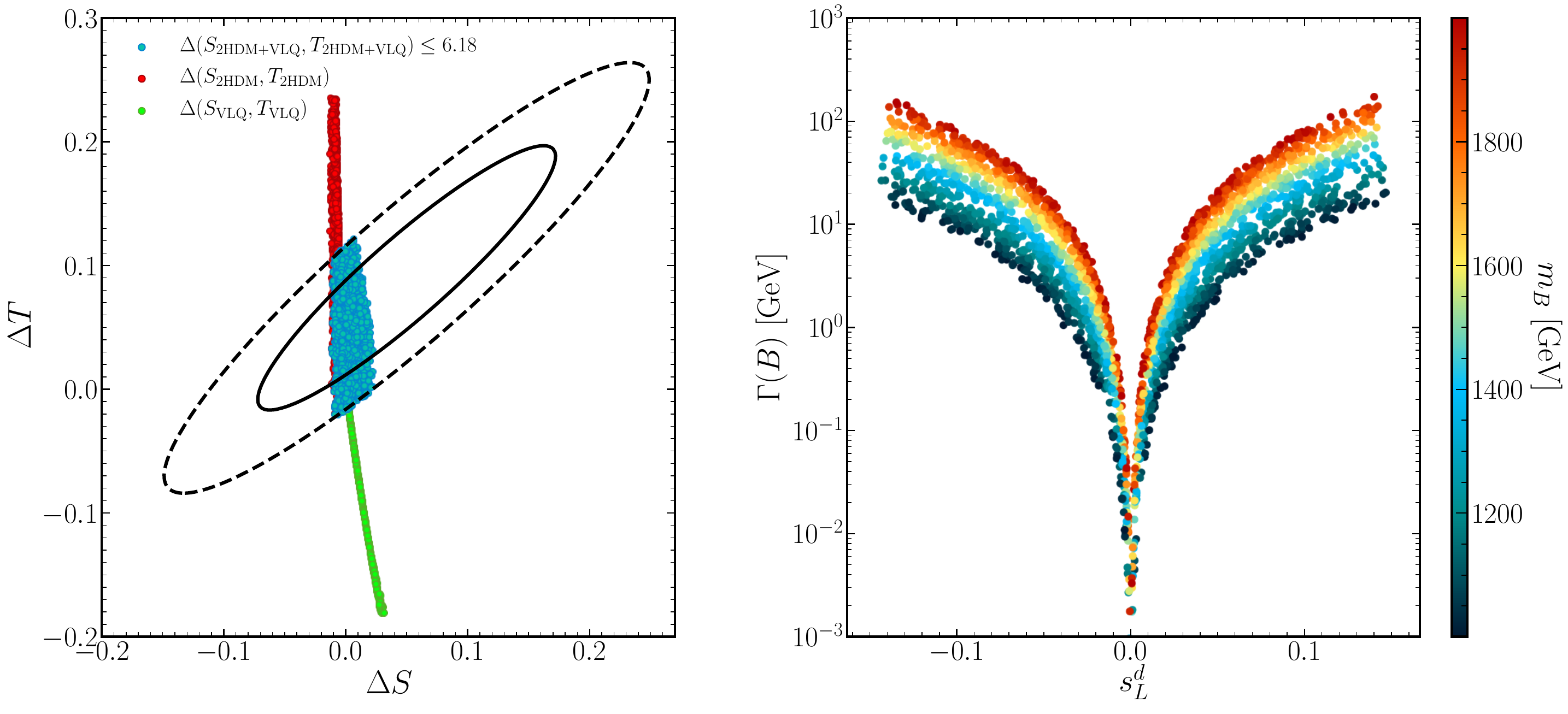}
	\caption{(Left) Scatter plots of randomly generated points superimposed onto the 
		fit limits in the $ (\Delta S, \Delta T)$ plane from EWPO data at 95\% CL
		with a correlation of 92\%. Here,  we illustrate the 2HDM and VLQ contributions separately and also the total one. (Right) The $B$ width $\Gamma(B)(\equiv \Gamma_B)$ as a function of $s_L^d\equiv\sin\theta_L$ with $m_B$ indicated as a color gauge.}
	\label{fig1}
\end{figure}
In Fig.~\ref{fig1}, we analyze the 2HDM parameters, which include the Higgs boson masses, $\tan\beta$, $\sin(\beta-\alpha)$, and $m_{12}$, along with the mass of the singlet VLB, $m_B$, and the fermionic mixing angle, $\sin\theta_L$, as detailed in Tab.~\ref{table1}.Unlike the analysis in Ref. \cite{Arhrib:2024tzm}, which focused on VLTs starting at $m_T = 650$ GeV, our study begins with $m_B = 1$ TeV (see Fig. \ref{fig_1}). This choice is due to the absence of recent LHC limits for VLBs below 1 TeV, whereas recent limits for VLTs start at 600 GeV \cite{CMS:2023agg}.

In the left panel of Fig.~\ref{fig1}, we detail our findings on how the 2HDM scalars and the single VLQ in this 2HDM+VLQ scenario influence the $S$ and $T$ parameters. We analyze these effects both separately and in combination.

The contributions to the $S$ parameter from the VLQ and 2HDM sectors, labeled $S_{\rm VLQ}$ and $S_{\rm 2HDM}$ respectively, are relatively minor. However, the impacts on the $T$ parameter, indicated as $T_{\rm VLQ}$ and $T_{\rm 2HDM}$, can be substantial and frequently exhibit opposing signs. This opposition allows for significant cancellations, particularly because $T_{\rm 2HDM}$ can be either positive or negative, whereas $T_{\rm VLQ}$ is nearly always negative. A pronounced negative $T_{\rm VLQ}$ can occur with high values of $\sin\theta_L$, and a large positive $T_{\rm 2HDM}$ can emerge in the 2HDM when there is considerable mass splitting among the heavy Higgs bosons. Notably, within this model, the constraints from the $S$ and $T$ parameters are stricter than those from $R_b$ limits \cite{Aguilar-Saavedra:2013qpa}.

The right panel of Fig.~\ref{fig1} projects our analysis onto the $(\sin\theta_L, \Gamma(B))$ plane\footnote{Here, $\Gamma(B)$, also referred to as $\Gamma_B$, represents the total width of the VLB.}. This illustrates that the mixing angle $|\sin\theta_L|$ generally remains below 0.15 across the entire $m_B$ range. For low mixing ($|\sin\theta_L|<0.1$), the total width is relatively small, often just a few GeV or less. Conversely, for larger mixing ($|\sin\theta_L| \approx 0.1-0.15$) and high $m_B$ values, the total width can vary between 10 and 100 GeV. These $\Gamma(B)$ values are likely comparable to the experimental resolution for reconstructed VLBs, implying that the width of the VLB could be constrained to a maximum of 10\% of its mass.

\begin{figure}[h!]
	\centering
	\includegraphics[width=0.45\textwidth,height=0.45\textwidth]{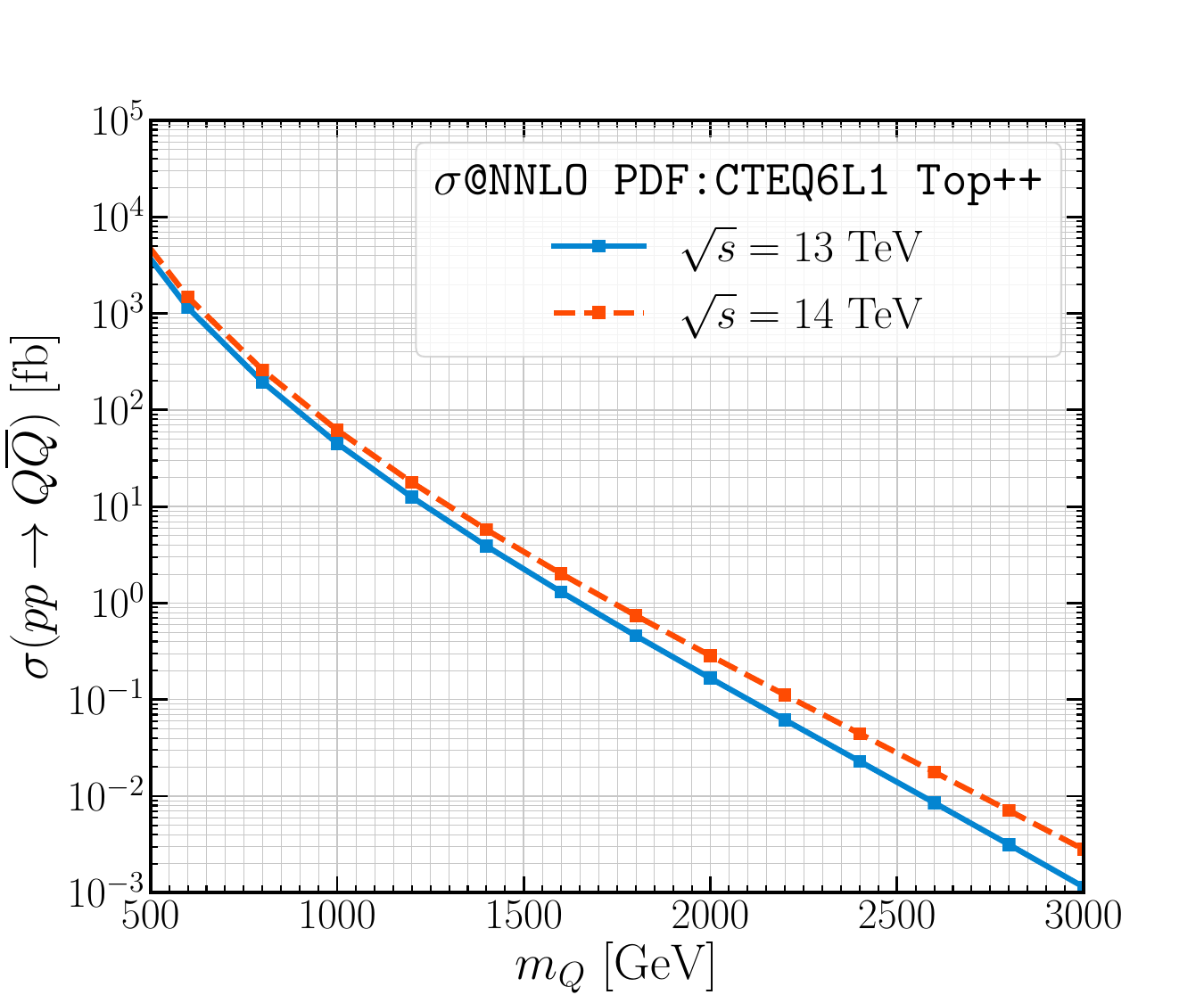}
	\caption{Pair production cross-section $\sigma(pp\to Q\bar{Q})$, computed at NNLO using 
		\texttt{Top++}\cite{Czakon:2011xx} for center-of-mass energies $\sqrt{s}$ = 13 TeV (blue) and  $\sqrt{s}$ = 14 TeV (orange), respectively.}	
	\label{xs_pair}
\end{figure}

This detailed analysis highlights the critical impact of adding extra degrees of freedom from both the VLQ sector and the 2HDM. The integration of these elements provides a broad spectrum of solutions, accommodating lighter $m_B$ masses (around 1 TeV) and larger mixing angles $|\sin\theta_L|$.  This could substantially enhance $pp\to B\overline{B}$ production rates at the LHC (see Fig.~\ref{xs_pair}) and increase the likelihood of $B$ decays into heavy (pseudo)scalar Higgs states from the 2HDM sector, as we shall discuss.

Fig.~\ref{xs_pair}, shows the pair production cross-section\footnote{It is important to note that the pair production of VLQs is model-independent and depends solely on their mass.} $\sigma(pp\to Q\bar{Q})$ at next-to-next-to-leading order (NNLO) using the \\\texttt{CTEQ6L1} \cite{Pumplin:2002vw} parton density sets, as a function of their mass $m_Q$. The curves are depicted in blue and orange, corresponding to center of mass energies of 13 TeV and 14 TeV, respectively.

Advancing our discussion to the Branching Ratios (${\cal BR}$) associated with the $B$ state, we find that the scenario within the SM, supplemented by an additional singlet bottom quark, closely mirrors the dynamics outlined for the singlet VLT scenario discussed in Ref.\cite{Arhrib:2024tzm}. The distribution of ${\cal BR}$s among $B\to W^-t$, $B\to Zb$, and $B\to hb$ remains approximately at 50\%, 25\%, and 25\%, respectively \cite{Aguilar-Saavedra:2013qpa}, similar to the VLT case. Notably, these distributions exhibit limited sensitivity to variations in the $\sin\theta_L$ mixing angle and adhere to a sum rule ensuring their collective contribution to the $B$ state decay processes\footnote{It is important to note that in this study, we focus solely on the on-shell decays of $B$, i.e., $m_B > m_{\mathrm{Higgs}}$.} into SM final states respects

\begin{eqnarray}
{\cal BR}(B\to {\rm SM}) &=& {\cal BR}(B\to W^-t)+ {\cal BR}(B\to Zb)\nonumber \\&&+ {\cal BR}(B\to hb)=1.
\label{eq:sumrule1} 
\end{eqnarray}

The introduction of additional decay pathways through (pseudo)scalar channels such as $B\to H^-t$, $B\to Ab$, and $B\to Hb$ results in a significant shift. This evolution diversifies the decay mechanisms and modifies the constraints previously established by direct $B$ searches at the LHC, leading to an updated sum rule:
\begin{eqnarray}
&&{\cal BR}(B\to {\rm SM}) + {\cal BR}(B\to {\rm non~SM}) =1,\nonumber \\
&& {\cal BR}(B\to {\rm non~SM}) = {\cal BR}(B\to H^-t)+ {\cal BR}(B\to Ab)\nonumber \\&&\hspace{3cm}+ {\cal BR}(B\to Hb).
\label{eq:sumrule2} 
\end{eqnarray}

\begin{figure*}[h!]
	\centering
	\includegraphics[width=0.9\textwidth,height=0.40\textwidth]{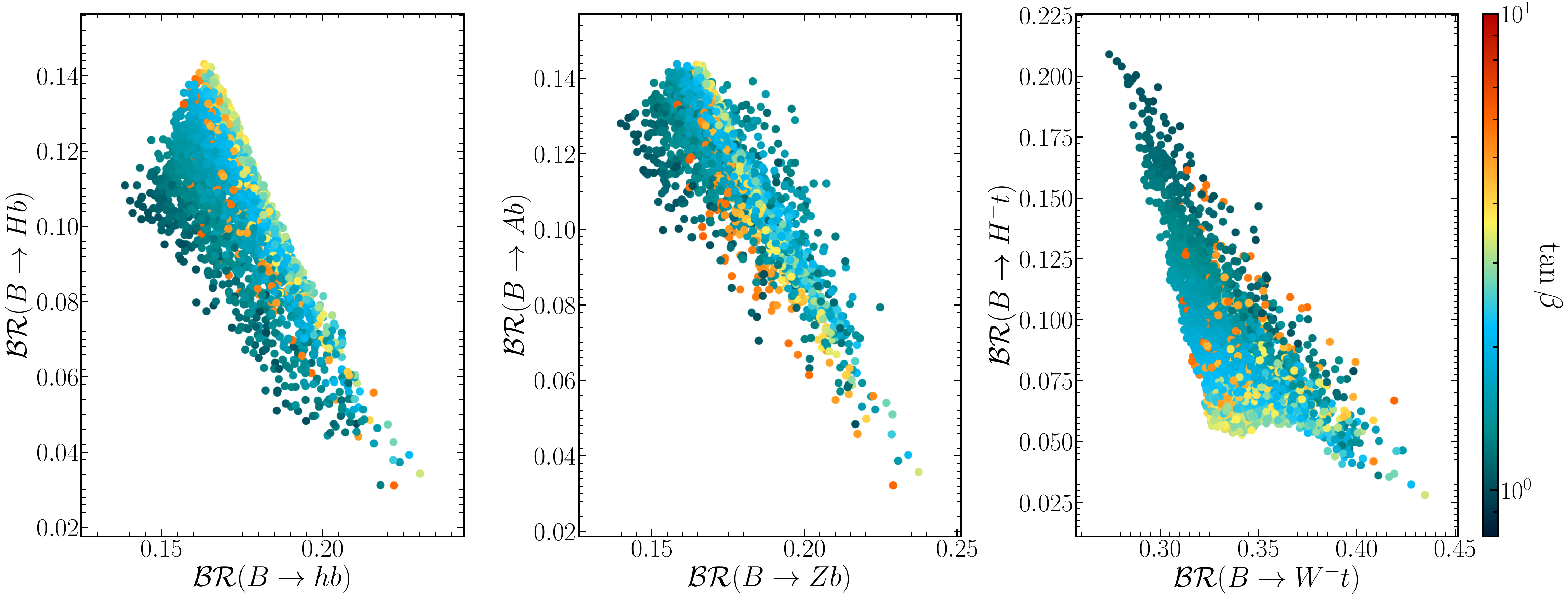}
	\caption{The correlation between ${\cal BR}(B\to hb)$ and ${\cal BR}(B\to Hb)$ (left), ${\cal BR}(B\to Zb)$ and ${\cal BR}(B\to Ab)$ (middle) as well as  ${\cal BR}(B\to W^-t)$ and ${\cal BR}(B\to H^-t)$ (right) with $\tan\beta$ indicated in the color gauge.}	
	\label{fig2}
\end{figure*}

In Fig.~\ref{fig2}, we illustrate the correlation between the decay processes of the $B$ state into SM particles and novel pathways, organized as follows: $B\to hb, Hb$ (left), $B\to Zb, Ab$ (middle), and $B\to W^-t, H^-t$ (right). It is observed that when non-SM decay channels become kinematically accessible, the ${\cal BR}(B\to H^-t)$ begins to rival the ${\cal BR}(B\to W^-t)$, especially for low $\tan\beta \sim 1$. A similar competitive relationship is noted between ${\cal BR}(B\to Hb)$ and ${\cal BR}(B\to Ab)$ for medium $\tan\beta$. {In turn, both $H$ and $A$ predominantly decay into $t\bar{t}$ (approaching 100\%), leading to the final state of $2W + 3b$ from the decay of a single $B$ quark. In the case of pair production of $B$, the final state would be $4W + 6b$, offering a distinct signature that can be probed at the LHC \cite{Dermisek:2020gbr}.}

\begin{figure}[h!]
	\centering
	\includegraphics[width=0.85\textwidth,height=0.4\textwidth]{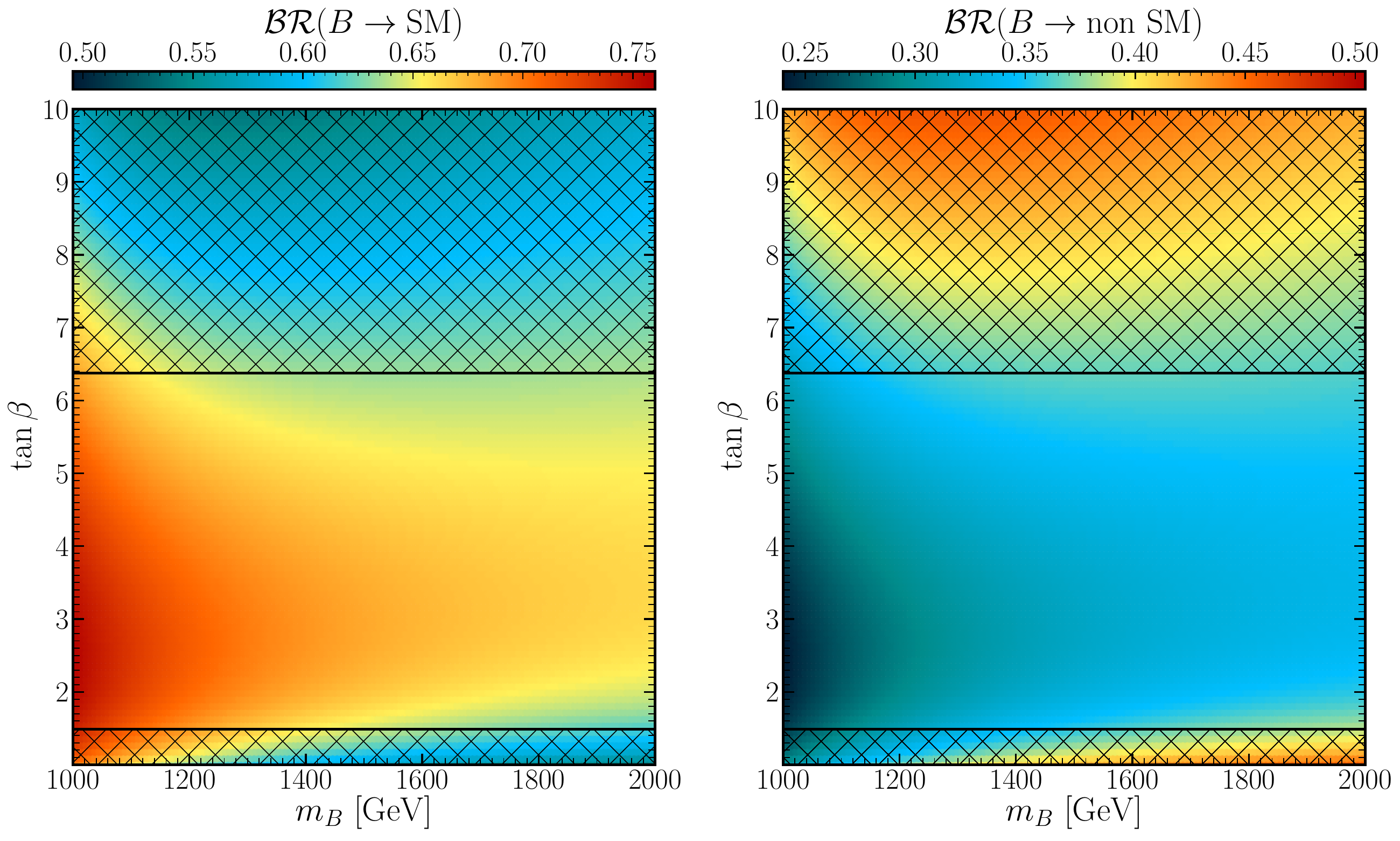}
	\caption{The ${\cal BR}(B\to$ SM) (left) and ${\cal BR}(B\to$ non SM)  (right) mapped onto the $(m_B, \tan\beta)$ plane, with $\sin\theta_L=0.045$, $\sin(\beta-\alpha)=1$, $m_h=125$ GeV, $m_H=585$, $m_A=582$ GeV, and $m_{H^\pm}=650$ GeV (recall that $m_{12}^2=m_A^2\tan\beta/(1+\tan^2\beta)$).  Here, the shaded areas are excluded by \texttt{HiggsBounds} ($H^+\to t\bar{b}$ \cite{ATLAS:2021upq} for $\tan\beta<2$ and $H\to\tau\tau$ \cite{ATLAS:2020zms} for $\tan\beta>6$). All other constraints ($S$, $T$, \texttt{HiggsSignals} and theoretical ones) are also checked.}	
	\label{fig3}
\end{figure}
	\begin{table}[h!]
	\begin{center}
		\setlength{\tabcolsep}{45pt}
		\renewcommand{\arraystretch}{1}
		\begin{adjustbox}{max width=\textwidth}		
			\begin{tabular}{lcc}
				\toprule\toprule
				Parameters &       BP$_1$ &       BP$_2$ \\
				\toprule

				\multicolumn{3}{c}{2HDM+VLQ inputs. The masses are in GeV.} \\\toprule
				$m_h$   &   125&   125  \\
				$m_H$  &    789.60 &  681.89\\
				$m_A$   &766.47 &  679.47\\
				$m_{H\pm}$   &750.39 &  667.01  \\
				$\tan\beta$ &    2.80 &    4.92 \\
				
				$m_B$      &1182.82 & 1897.84\\
				$\sin(\theta^d)_L$    &  0.035 &    0.016\\
				
				\toprule
				
				\multicolumn{3}{c}{$\mathcal{BR}(T\to {XY})$ in \%} \\\toprule
				${\cal BR}(B\to W^+t)$  &   39.18 & 33.14 \\
				${\cal BR}(B\to W^+T)$  &   - &   - \\
				${\cal BR}(B\to Zb)$  &   20.87 & 16.98\\
				${\cal BR}(B\to hb)$  &   20.41 & 16.84   \\
				${\cal BR}(B\to Hb)$  &   6.42 & 12.88    \\
				${\cal BR}(B\to Ab)$  &  7.02 & 12.91\\
				${\cal BR}(B\to H^+t)$ &  6.10 &  7.26 \\
				${\cal BR}(B\to H^+T)$ &   - &  -   \\				
				\toprule
				\multicolumn{3}{c}{Total decay width in GeV.} \\\toprule
				$\Gamma(B)$ &  1.59 &    1.62 \\
				\toprule
				
				\multicolumn{3}{c}{Observables} \\\toprule
				$T_{\mathrm{2HDM}}$  &  0.0112 &    0.0033 \\
				$T_{\mathrm{VLQ}}$  &  -0.0118 &   -0.0029 \\
				$S_{\mathrm{2HDM}}$ &  0.0018 &    0.0010\\
				$S_{\mathrm{VLQ}}$ &    0.0016 &    0.0004   \\	
				$\Delta\chi^2(S_{\mathrm{2HDM+VLQ}},T_{\mathrm{2HDM+VLQ}})$ &    4.08 &    3.75\\\toprule
				$\chi^2{(h_{125})}\equiv \chi^2_{\texttt{HiggsSignals}}$ &  158.61 &  158.70  \\	
				
				\bottomrule\bottomrule
				
			\end{tabular}
		\end{adjustbox}
	\end{center}
	\caption{The full description of our BPs for the $(B)$ singlet case.}\label{Bp1}
\end{table}
In Fig.~\ref{fig3}, we present the cumulative ${\cal BR}$s of the $B$ state into SM particles (left panel) and non-SM particles (right panel), plotted against the singlet bottom mass $m_B$ and the parameter $\tan\beta$. It is evident that for high values of $\tan\beta$, the decay rates into non-SM particles increase significantly, reaching up to 50\%\footnote{{Our analysis focuses on the on-shell decays of $B$ quarks, i.e. $m_B > m_{\mathrm{Higgs}}$, with Higgs masses kept below 1 TeV. This choice restricts $\tan\beta$ to values below approximately 6, as constrained by LHC $H\to \tau\tau$ searches \cite{ATLAS:2020zms}. For Higgs masses exceeding 1 TeV, larger $\tan\beta$ values could be allowed, potentially increasing the ${\cal BR}$ for non-SM decays up to 50\%.}} (38\% within the permissible region). In contrast, at medium and low $\tan\beta$ values, irrespective of $m_B$, the scenario reverses, with decays into SM particles becoming fully predominant.
	
	In this context, we propose three Benchmark Points (BPs) for investigating the Type-II 2HDM extended by a singlet VLQ carrying the bottom electric charge. The proposed BPs are as follows\footnote{If these conditions are not fully satisfied, we will proceed with only two BPs instead of the proposed three.}:
	
	\begin{itemize}
		\item[ i)] BP$_1$: where ${\cal BR}(B\to$ SM) is similar to ${\cal BR}(B\to$ non-SM),
		\item[ii)] BP$_2$: where ${\cal BR}(B\to$ SM) is relatively small, making ${\cal BR}(B\to$ non-SM) substantial,
		\item[iii)] BP$_3$: where ${\cal BR}(B\to$ SM) is relatively large, making ${\cal BR}(B\to$ non-SM) marginal.
	\end{itemize}

Input parameters for these BPs are given in Tab.~\ref{Bp1}.

\subsection{2HDM with $(TB)$ doublet}

We now examine the scenario involving the $(TB)$ doublet. In the SM extended by this VLQ multiplet, both mixing angles in the up-and down-type quark sectors enter the phenomenology of the model\footnote{Hereafter, we will denote the $d$ and $u$ superscripts in the mixing angles as $b$ and $t$, respectively.}. For given $\theta_R^b$, $\theta_R^t$, and $m_B$, the relationship between the mass eigenstates and the mixing angles is expressed as \cite{Aguilar-Saavedra:2013qpa}:
\begin{eqnarray}
&&m_T^2= (m_B^2 \cos^2\theta_R^{b}+m_b^2 \sin^2\theta_R^{b}-
m_t^2 \sin^2\theta_R^{t})/ \cos^2\theta_R^{t},
\label{tb-mass}
\end{eqnarray}

from which the $m_T$ value can be derived. Additionally, using Eqs.~(\ref{ec:angle1})--(\ref{ec:rel-angle1}), the left mixing angles $\theta_L^b$ and $\theta_L^t$ can also be calculated.

{For our analysis of the 2HDM+$(TB)$ doublet, we adopt the scan ranges provided in Tab.\ref{table1}. Fig.~\ref{fig4} illustrates the $S$ and $T$ parameters, distinguishing the individual contributions from the VLQ and the 2HDM. As shown in the left panel of Fig.\ref{fig4}, similar to the heavy bottom $(B)$ singlet case, the VLQ $(TB)$ doublet and 2HDM states can contribute with opposite signs to the oblique parameters. This feature can significantly influence the constraints on the 2HDM+VLQ parameter space derived from EWPOs. Moreover, as depicted in the right panel of the figure, within the SM extended by a VLQ (green points), the mass splitting between $B$ and $T$ must not exceed 25 GeV positively and 5 GeV negatively. However, introducing 2HDM states with VLQs expands this allowable mass splitting, reaching up to 40 GeV positively and 10 GeV negatively.

	\begin{figure}[H]
		\centering
		\includegraphics[width=0.85\textwidth,height=0.4\textwidth]{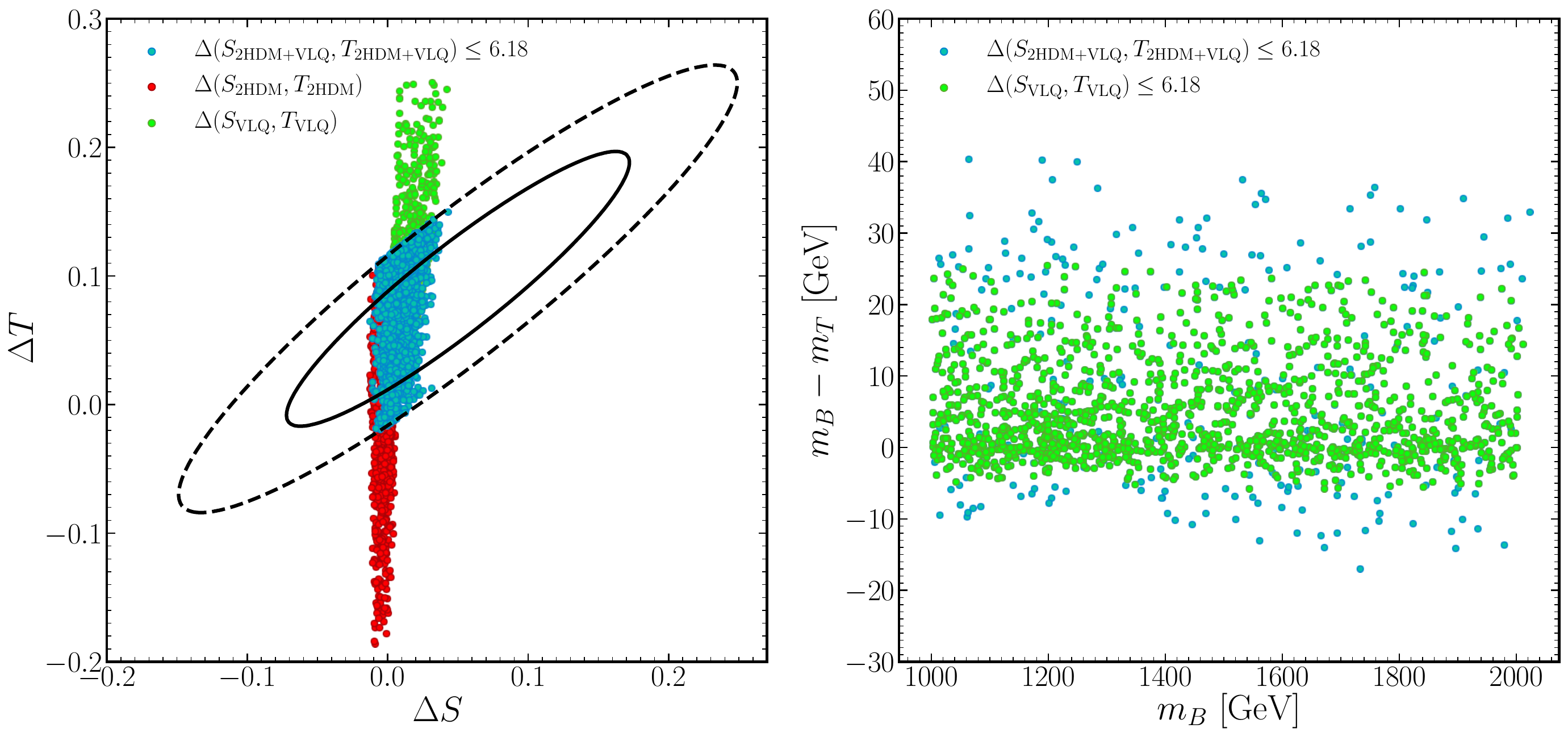} 	
		\caption{(Left) Scatter plots of randomly generated points superimposed onto the 
			fit limits in the $ (\Delta S, \Delta T)$ plane from EWPO data at 95\% CL
			with a correlation of 92\%. Here,  we illustrate the 2HDM and VLQ contributions separately and also the total one. 
			(Right) The same points are mapped onto the $(m_B,\delta)$ plane, where $\delta$ is the mass difference between $B$ and $T$. Here, we only present the VLQ contribution and the total one.  Further, all constraints have been taken into account.} 	
		\label{fig4}
	\end{figure}
	
		\begin{figure}[H]
		\centering
		\includegraphics[width=0.85\textwidth,height=0.4\textwidth]{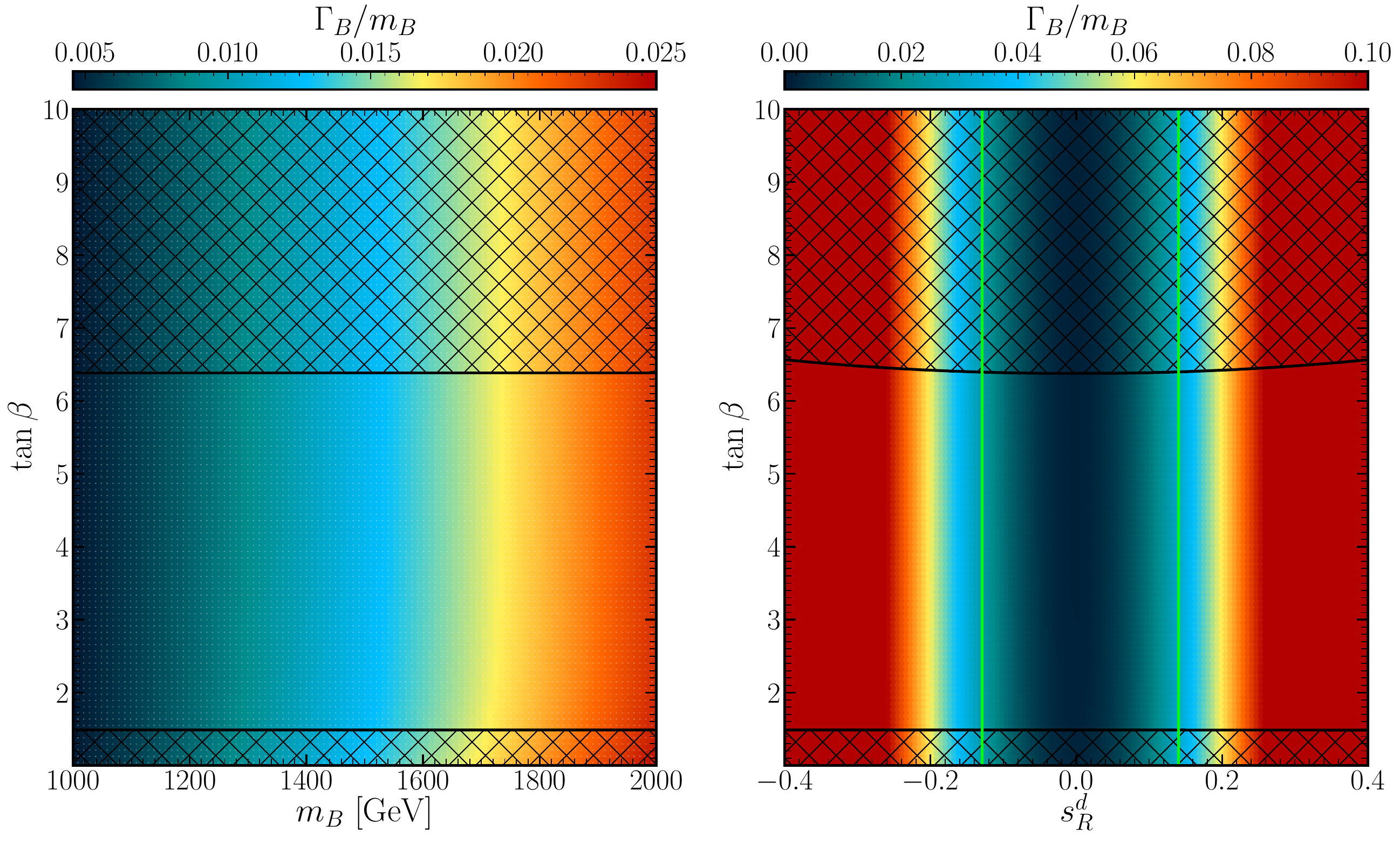}
		\caption{The $\Gamma(B)/m_B$ ratio $(\Gamma(B)$ mapped over the $(m_B,\tan\beta)$ plane (left) and $(s^d_R\equiv\sin\theta^d_R,\tan\beta)$ plane (right), with $\sin\theta^u_R=0.042$ in the left panel and $m_B=1600$ GeV in the right panel (the 2HDM parameters are the same as in Fig. \ref{fig3}). Here, the shaded areas are excluded by \texttt{HiggsBounds}. The regions between the vertical lime green lines are allowed by the $S$, $T$ parameter constraints, all other constraints  (\texttt{HiggsSignals} and theoretical ones) are also checked.}	
		\label{fig5}
	\end{figure}
	
	In the scenario where the SM is extended by a VLQ $(TB)$ doublet, the $B$ state can decay through various modes: $B \rightarrow {W^-t, Zb, bh}$ \cite{Aguilar-Saavedra:2013qpa}. Notably, if the mixing angle $\theta^b_R$ is zero ($\theta^b_R=0$), the $bBZ$ and $bBh$ couplings vanish, resulting in $B \rightarrow W^-t$ being the dominant decay mode. When the mixing is non-zero, these decay channels become viable, with their branching ratios (${\cal BR}$s) depending on the mixing angles and VLQ masses. The addition of extra Higgs particles introduces new decay pathways: $B \rightarrow {Ab, Hb, H^-t}$\footnote{It is important to note that the mass splitting constraints between $B$ and $T$, as dictated by the $S$ and $T$ parameters, limit the potential for VLQ decays into each other, such as $B \rightarrow W^-/H^- T$. Hence, we will focus solely on the kinematically allowed $1\to2$ channels (on-shell decays).} \cite{Aguilar-Saavedra:2017giu, Dermisek:2019vkc}.
	
	To further investigate the phenomenology of this BSM setup, we conduct a comprehensive scan of both the 2HDM and VLQ $(TB)$ doublet parameters as outlined in Tab.~\ref{table1}, ensuring compliance with all theoretical and experimental constraints. Fig.~\ref{fig5} depicts the total width of the $B$ state as a function of its mass $m_B$ (left panel) and the down-quark sector mixing angle $s^d_R$ (right panel), with both dimensions correlated to $m_B$ and showing insensitivity to $\tan\beta$. The analysis reveals that $\Gamma(B)/m_B$ increases significantly with $m_B$, and, as indicated in the right panel, this trend is particularly pronounced for large mixing values ($|s^d_R|\ge 0.2$), which exceed the permissible bounds set by the $S$ and $T$ parameters (green line).

		\begin{figure}[H]
		\centering
		\includegraphics[width=0.85\textwidth,height=0.4\textwidth]{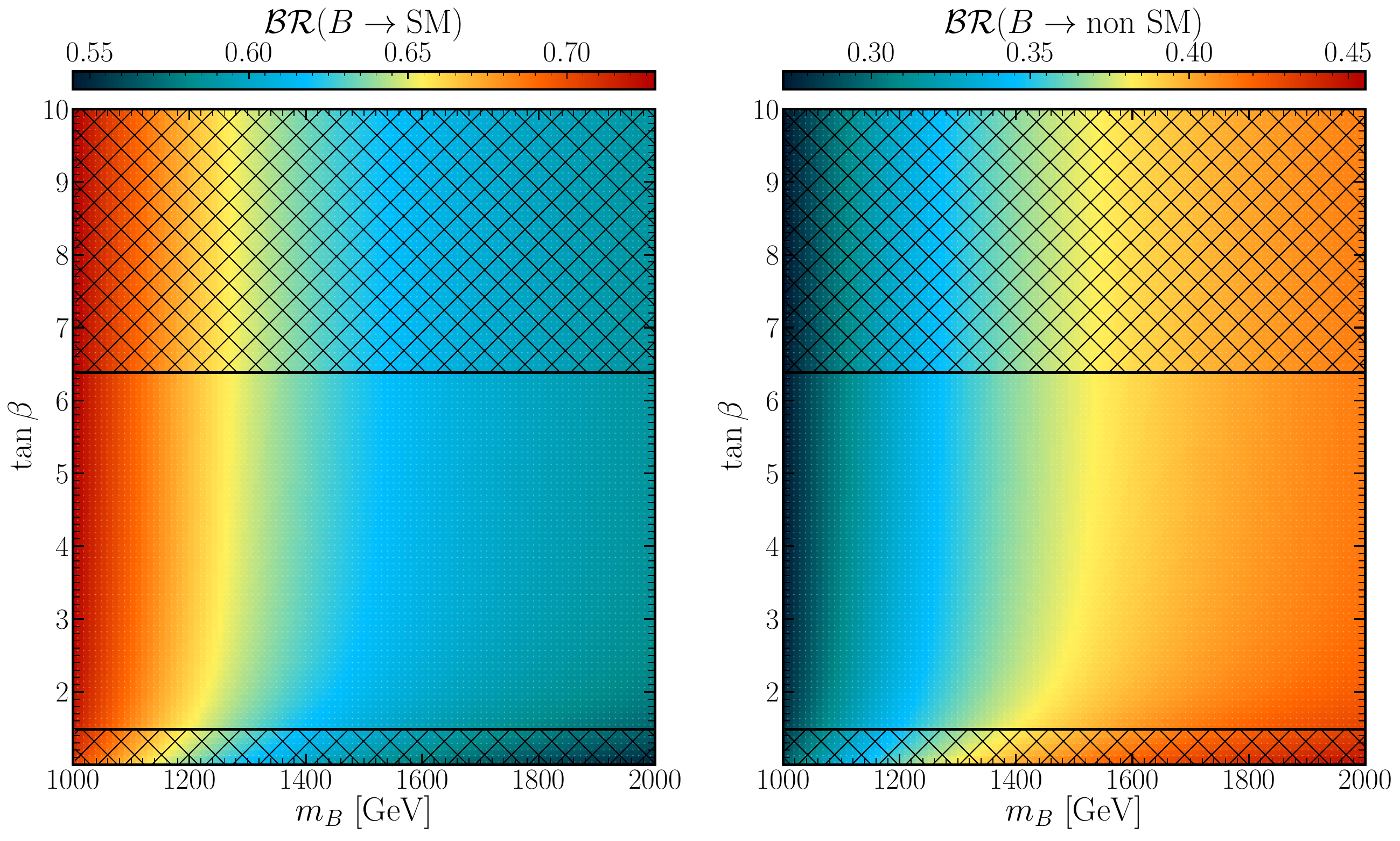}
		\caption{The ${\cal BR}(B\to$ SM) (left) and ${\cal BR}(B\to$ non SM)  (right) mapped onto the $(m_B, \tan\beta)$ plane, with the same description as in Fig.~\ref{fig5} (left).}	
		\label{fig6}
	\end{figure}

	\begin{figure}[H]
		\centering
		\includegraphics[width=0.9\textwidth,height=0.40\textwidth]{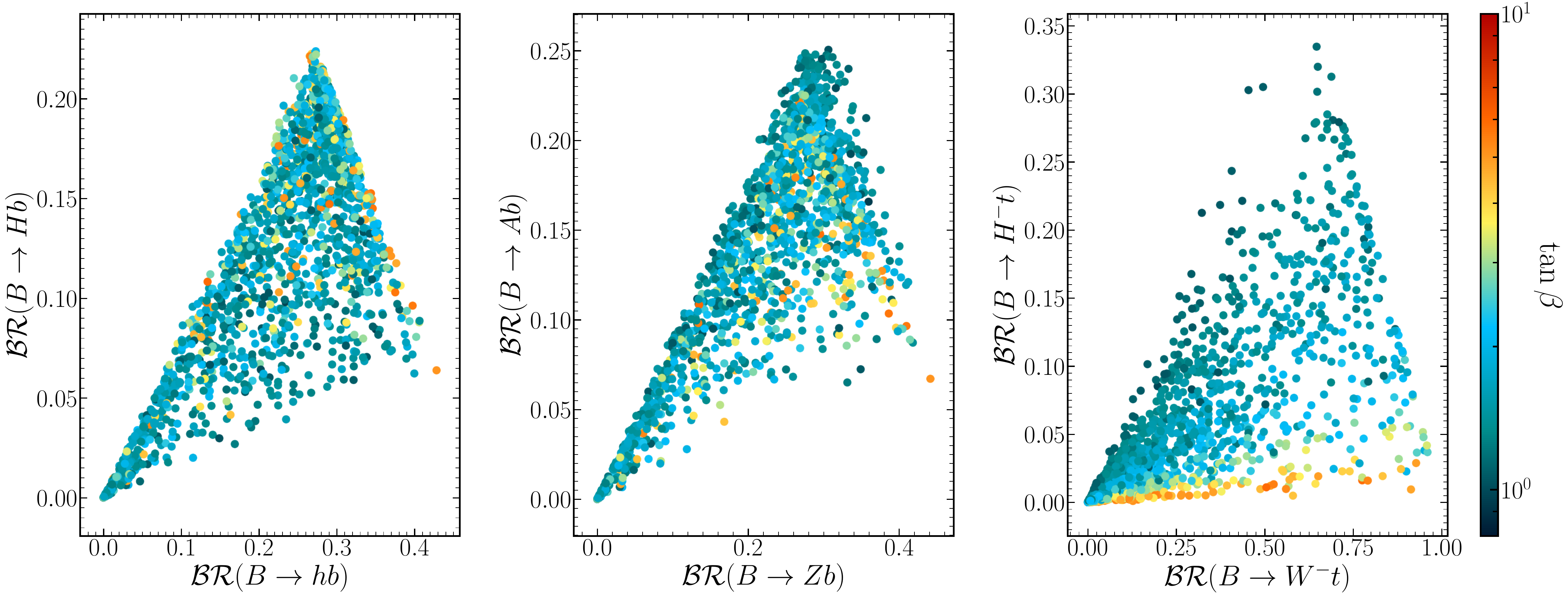} 	
		\caption{The correlation between ${\cal BR}(B\to hb)$ and ${\cal BR}(B\to Hb)$  (left),   
			${\cal BR}(B\to Zb)$ and ${\cal BR}(B\to Ab)$ (middle) as well as  ${\cal BR}(B\to W^-t)$ and ${\cal BR}(B\to H^-t)$ (right) with $\tan\beta$ indicated in the color gauge.}	
		\label{fig7}
	\end{figure}
	
	Fig.~\ref{fig6}  examines the decay patterns into SM and non-SM particles, the latter involving additional Higgs states such as $B \rightarrow H^- t$, $B \rightarrow Ab$, and $B \rightarrow Hb$\footnote{In principle, the definitions of these two ${\cal BR}$s should include $B \rightarrow W^-T$ and $B \rightarrow H^-T$ decays. However, since both $W^-$ and $H^-$ are off-shell, they are always small, so we will not discuss these here (recall the degeneracy $m_B\sim m_T$).}. This analysis shows that non-SM decay channels of the $B$ state can account for up to 40\% of decays at higher $m_B$ values, with minimal dependence on $\tan\beta$.

	Fig.~\ref{fig7}, further explores the correlation between the individual SM and non-SM decay channels of the $B$ state. This analysis indicates that non-SM decay channels generally have lower ${\cal BR}$s compared to their SM counterparts in both neutral and charged current scenarios. Specifically, the ${\cal BR}$ for $B \rightarrow Zb$ and $B \rightarrow hb$ can reach approximately 45\%, while for $B \rightarrow Ab$ and $B \rightarrow Hb$, it can reach around 25\% at most.
	\begin{table}[H]
	\begin{center}
		\setlength{\tabcolsep}{45pt}
		\renewcommand{\arraystretch}{0.8}
		\begin{adjustbox}{max width=\textwidth}		
			\begin{tabular}{lcc}
				\toprule\toprule
				Parameters &       BP$_1$ &       BP$_2$ \\
				\toprule

				\multicolumn{3}{c}{2HDM+VLQ inputs. The masses are in GeV.} \\\toprule
				$m_h$   &   125&   125  \\
				$m_H$  &    823.71 &  815.62\\
				$m_A$   &814.69 &  707.44\\
				$m_{H\pm}$   &870.31 &  706.66  \\
				$\tan\beta$ &   3.43 &    1.69 \\
				$m_T$      & 1160.95 & 1901.09\\
				$m_B$      & 1172.58 & 1918.24\\
				
				$\sin(\theta^u)_L$    &  0.0101 &   -0.0033\\
				$\sin(\theta^d)_L$    & -0.0006 &    0.0003\\
				$\sin(\theta^u)_R$    &  0.0676 &   -0.0365\\
				$\sin(\theta^d)_R$    & -0.1553 &    0.1382 \\
				\toprule
				\multicolumn{3}{c}{$\mathcal{BR}(T\to {XY})$ in \%} \\\toprule
				${\cal BR}(B\to W^+t)$  &   12.66 &  3.83  \\
				${\cal BR}(B\to W^+T)$  &   - &   - \\
				${\cal BR}(B\to Zb)$  &   34.85 & 27.92 \\
				${\cal BR}(B\to hb)$  &   34.07 & 27.69   \\
				${\cal BR}(B\to Hb)$  &   8.94 & 18.74   \\
				${\cal BR}(B\to Ab)$  &  9.33 & 20.85 \\
				${\cal BR}(B\to H^+t)$ &  0.15 &  0.97 \\
				${\cal BR}(B\to H^+T)$ &   - &  -   \\				
				\toprule
				\multicolumn{3}{c}{Total decay width in GeV.} \\\toprule
				$\Gamma(B)$ &  42.49 &  137.69  \\
				\toprule
				\multicolumn{3}{c}{Observables} \\\toprule
				$T_{\mathrm{2HDM}}$  &   0.0462 &    0.0015 \\
				$T_{\mathrm{VLQ}}$  &  0.0964 &    0.0936\\
				$S_{\mathrm{2HDM}}$ &  -0.0031 &    0.0039\\
				$S_{\mathrm{VLQ}}$ &    0.0192 &    0.0165  \\	
				$\Delta\chi^2(S_{\mathrm{2HDM+VLQ}},T_{\mathrm{2HDM+VLQ}})$ &    2.20 &    0.30\\\toprule
				$\chi^2{(h_{125})}\equiv \chi^2_{\texttt{HiggsSignals}}$ & 157.08 &  157.37 \\	
				
				\bottomrule\bottomrule
				
			\end{tabular}
		\end{adjustbox}
	\end{center}
	\caption{The full description of our BPs for the $(TB)$ doublet case.}\label{Bp3}
\end{table}		
	Moreover, the analysis reveals a notable variation in the charged current decays: the non-SM decay $B \rightarrow H^-t$ can achieve a maximum ${\cal BR}$ of 34\% at low $\tan\beta$ values. In contrast, the SM decay $B \rightarrow W^-t$ can reach up to 100\% at medium $\tan\beta$ levels. This difference underscores the influence of $\tan\beta$ on the decay behavior of the $B$ state, with the couplings $bBH$ and $bBA$ showing a slight increase proportional to $\tan\beta$ at medium levels of this parameter. A similar scenario involving the $TB$ doublet has been discussed in detail in Ref. \cite{Aguilar-Saavedra:2017giu}, where the dependence of these decays on $\tan\beta$ was also explored.
	
	Following the methodology of the previous section, we conclude this part by proposing two BPs in Tab.~\ref{Bp3} for experimental investigation. These BPs are designed with varying $B$ quark masses and widths ($\Gamma_B$) to cover a range of potential detection scenarios.


}	
\subsection{2HDM with $(BY)$ doublet}
\begin{figure}[H]
	\centering
	\includegraphics[width=0.85\textwidth,height=0.4\textwidth]{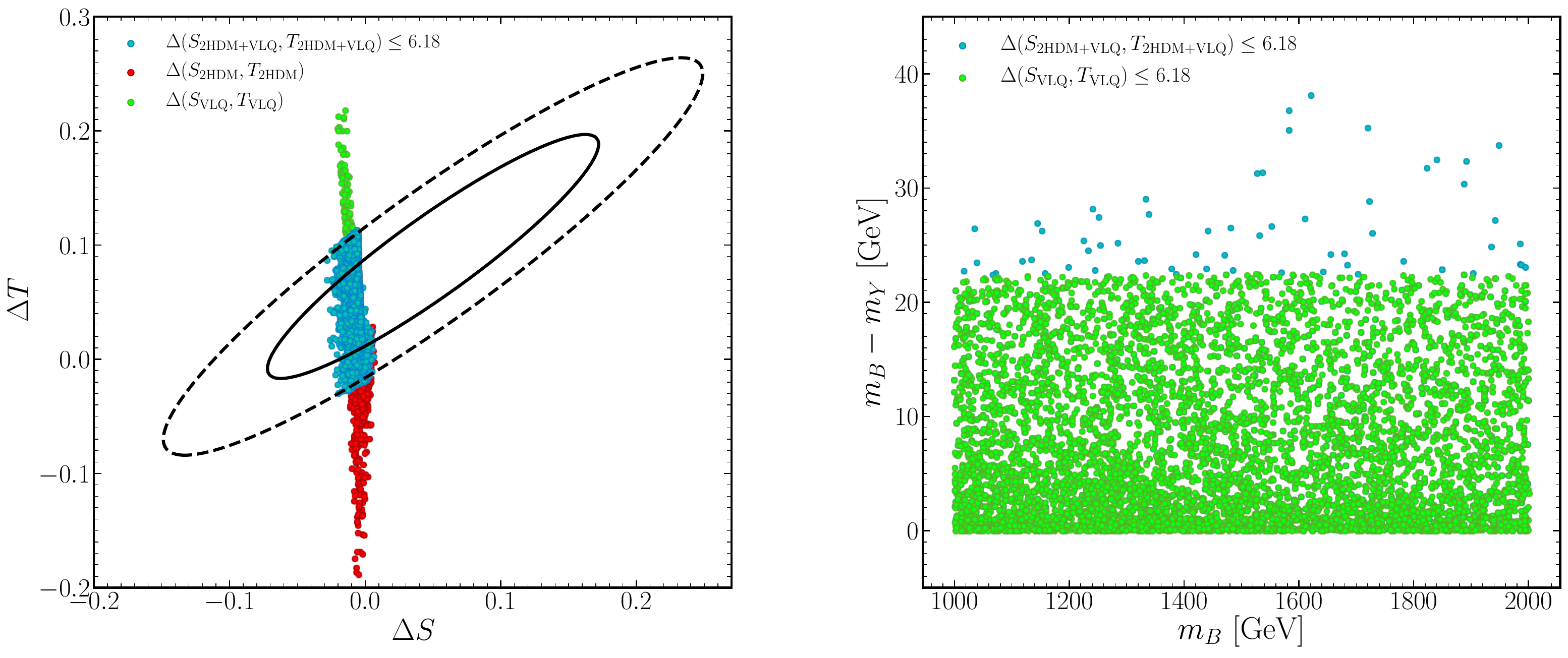}
	\caption{ (Left) Scatter plots of randomly generated points superimposed  onto the 
		fit limits in the $ (\Delta S, \Delta T)$ plane from EWPO data at 95\% CL with a correlation of 92\%. Here,  we illustrate the 2HDM and VLQ contributions separately and also the total one.
		(Right) The same points are here mapped onto the $(m_B,\delta)$ plane, where $\delta$ is the mass difference between $B$ and $Y$. Here, we only present the VLQ contribution and the total one. Further, all constraints have been taken into account.}
	\label{fig8}
\end{figure}

In the scenario where the SM is extended with a $(BY)$ doublet, the VLQ structure is fully characterized by the $\theta_R^b$ mixing angle and the new bottom mass $m_B$. For a given $\theta_R^b$ value, $\theta_L^b$ is calculated using Eq.~(\ref{ec:rel-angle1}). The mass of the new VLQ with an exotic EM charge of $+5/3$, referred to as the $Y$ state, is determined by the mixing angle $m_B$ and $m_b$ as per \cite{Aguilar-Saavedra:2013qpa}:
\begin{equation}
m_Y^2=m_B^2\cos\theta_R^2+m_b^2\sin\theta_R^2.\label{eq_BY}
\end{equation}
This is independent (at tree level) from the additional parameters entering the 2HDM Higgs sector, however, the latter impinges on the viability of this BSM construct against EWPO data. 

\begin{figure}[H]
	\centering 	
	\includegraphics[width=0.85\textwidth,height=0.4\textwidth]{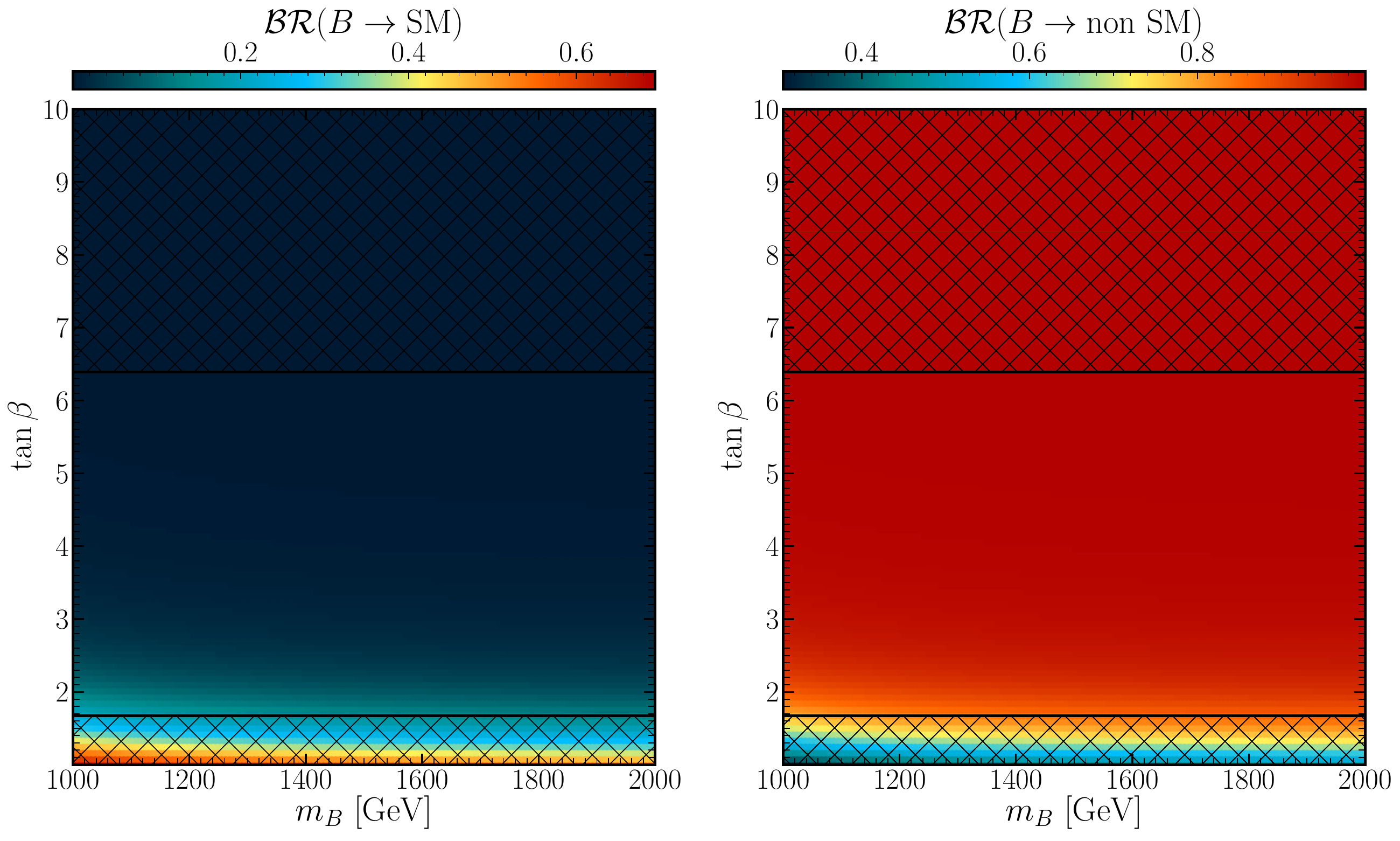}
	\caption{The ${\cal BR}(B\to$ SM) (left) and ${\cal BR}(B\to$ non SM)  (right) mapped onto the $(m_B, \tan\beta)$ plane. with $\sin\theta_R^u=0.057$ (the 2HDM parameters are the same as in Fig. \ref{fig3}). Here, the shaded areas are excluded by \texttt{HiggsBounds}, and all other constraints  ($S$, $T$, \texttt{HiggsSignals} and theoretical ones) are also checked.}	
	\label{fig9}
\end{figure}

\begin{figure}[H]
	\centering
	\includegraphics[width=0.85\textwidth,height=0.40\textwidth]{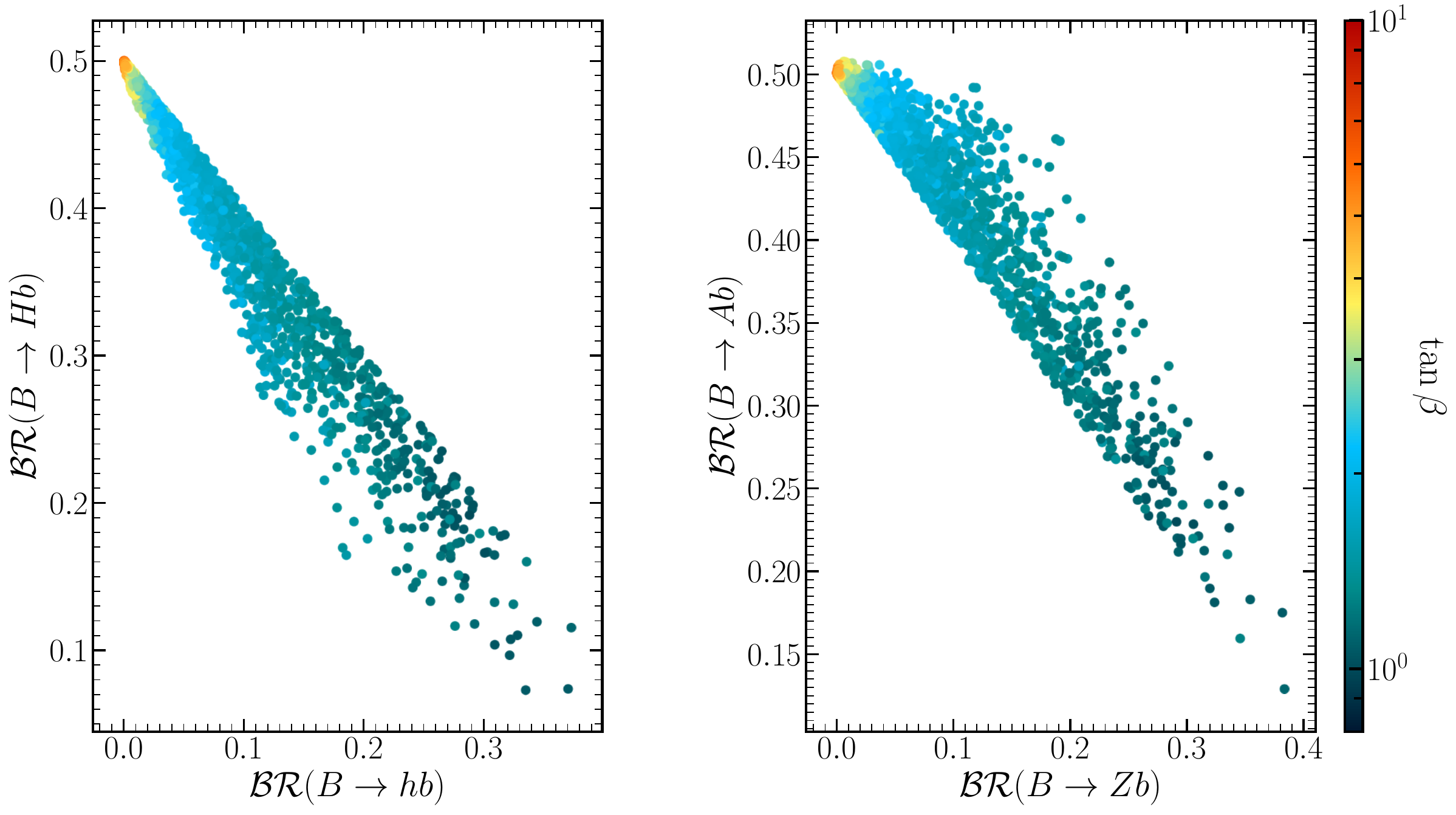}
	\caption{The correlation between ${\cal BR}(B\to hb)$ and ${\cal BR}(B\to Hb)$ (left),
		${\cal BR}(B\to Zb)$ and ${\cal BR}(B\to Ab)$ (right) with $\tan\beta$ as indicated in the color gauge.}
	\label{fig10}
\end{figure}

Building on the analysis in Tab.\ref{table1}, Fig.~\ref{fig8} (left) illustrates that while the combined (VLQ+2HDM) contributions fall within the $2\sigma$ confidence level of the $\Delta S$ and $\Delta T$ parameters (blue points), the individual contributions from the VLQ and 2HDM sectors frequently exceed the limits established by the EWPO data ellipses.

In the right panel of Fig.~\ref{fig8}, we show the allowable mass splitting $\delta = m_B - m_Y$ as constrained by EWPO data. For the VLQ-only structure, the mass splitting remains very small. However, in the full 2HDM+VLQ scenario, the splitting can be significantly larger, up to about 40 GeV, particularly for higher values of $m_B$.

Fig.~\ref{fig9} compares ${\cal BR}(B \rightarrow {\rm SM})$ to ${\cal BR}(B \rightarrow {\rm non~SM})$, mapped over $m_B$ and $\tan\beta$. Unlike the previous VLQ doublet scenario, here, the non-SM decays of the $B$ state can dominate, potentially reaching up to 100\% within the permissible parameter space, which notably expands with increasing $\tan\beta$. Interestingly, the decay behavior of the heavy bottom partner shows minimal sensitivity to the mass $m_B$.

We now discuss the sizes of the individual ${\cal BR}$s of $B$ decays. As usual, alongside the SM decays of the new bottom $B$ into $W^-t, Zb$, and $hb$, the non-SM decays $B \rightarrow Hb$ and $B \rightarrow Ab$ are also present. However, the $B \rightarrow H^-t$ channel is unavailable, as the associated coupling is identically zero, and hence, it is excluded from our discussion.	
\begin{table}[H]
	\begin{center}
		\setlength{\tabcolsep}{30pt}
		\renewcommand{\arraystretch}{1}
		\begin{adjustbox}{max width=\textwidth}		
			\begin{tabular}{lccc}
				\toprule\toprule
				Parameters &       BP$_1$ &       BP$_2$&       BP$_3$ \\
				\toprule

				\multicolumn{4}{c}{2HDM+VLQ inputs. The masses are in GeV.} \\\toprule
				$m_h$   &   125&   125  &   125 \\
				$m_H$  &    947.12 &  821.41 &  749.15\\
				$m_A$   & 845.34 &  721.89 &  654.05\\
				$m_{H\pm}$   & 916.27 &  745.21 &  646.01  \\
				$\tan\beta$ &    1.27 &    1.17 &    1.55  \\
				$m_B$      & 1030.63 & 1474.56 & 1809.87\\
				$m_Y$      & 1011.76 & 1459.47 & 1804.03\\
				$\sin(\theta^d)_L$    & -0.0009 &   -0.0005 &   -0.0002\\
				$\sin(\theta^d)_R$    & -0.1905 &   -0.1427 &   -0.0802\\
				\toprule
				\multicolumn{4}{c}{$\mathcal{BR}(T\to {XY})$ in \%} \\\toprule
				${\cal BR}(B\to W^+t)$  &  0.002 &  0.001 &  0.000 \\
				${\cal BR}(B\to Zb)$  &    43.32 & 25.15 &  9.74\\
				${\cal BR}(B\to hb)$  &   42.06 & 24.79 &  9.65  \\
				${\cal BR}(B\to Hb)$  &   2.69 & 22.60 & 38.38  \\
				${\cal BR}(B\to Ab)$  &   11.93 & 27.46 & 42.24\\
				${\cal BR}(B\to H^+t)$  &  0.00 &  0.00 &  0.00 \\
				\toprule
				\multicolumn{4}{c}{Total decay width in GeV.} \\\toprule
				$\Gamma(B)$ &  14.50 &   41.71 &   63.88  \\
				\toprule
				\multicolumn{4}{c}{Observables} \\\toprule
				$T_{\mathrm{2HDM}}$  &   -0.0389 &   -0.0316 &    0.0148 \\
				$T_{\mathrm{VLQ}}$  &  0.0761 &    0.0491 &    0.0073\\
				$S_{\mathrm{2HDM}}$ & -0.0011 &    0.0018 &    0.0042\\
				$S_{\mathrm{VLQ}}$ &    -0.0109 &   -0.0075 &   -0.0026  \\	
				$\Delta\chi^2(S_{\mathrm{2HDM+VLQ}},T_{\mathrm{2HDM+VLQ}})$ &     0.61 &    1.50 &    1.49\\\toprule
				$\chi^2{(h_{125})}\equiv \chi^2_{\texttt{HiggsSignals}}$ & 158.32 &  158.55 &  158.70 \\	
				
				\bottomrule\bottomrule
				
			\end{tabular}
		\end{adjustbox}
	\end{center}
	\caption{The full description of our BPs for the $(BY)$ doublet case.}\label{Bp4}
\end{table}

In Fig.~\ref{fig10}, we show the usual correlations between \\${\cal BR}(B \rightarrow hb)$ and ${\cal BR}(B \rightarrow Hb)$ (left) as well as ${\cal BR}(B \rightarrow Zb)$ and ${\cal BR}(B \rightarrow Ab)$ (right). At medium $\tan\beta$ values (with large $\tan\beta$ values ruled out by $H \rightarrow \tau\tau$ searches \cite{ATLAS:2020zms}), both ${\cal BR}(B \rightarrow Hb)$ and ${\cal BR}(B \rightarrow Ab)$ rise significantly, each reaching up to 50\% as their respective couplings increase proportionally to $\tan\beta$. This leads to a noticeable increase in ${\cal BR}(B \rightarrow Zb)$ and ${\cal BR}(B \rightarrow hb)$, making medium $\tan\beta$ an ideal range for investigating exotic $B$ decays within the established parameter space.

Conversely, at lower $\tan\beta$ values, ${\cal BR}(B \rightarrow hb)$ and ${\cal BR}(B \rightarrow Zb)$ receive a boost, each approaching approximately 38\%, which in turn mildly suppresses ${\cal BR}(B \rightarrow Hb)$ and ${\cal BR}(B \rightarrow Ab)$. This nuanced behavior across different $\tan\beta$ ranges highlights the complex interplay between these parameters and the decay modes of the $B$ state in this BSM framework.

In summary, at small $\tan\beta$ values, both \\${\cal BR}(B \rightarrow hb)$ and ${\cal BR}(B \rightarrow Zb)$ are somewhat enhanced, each nearing 38\%, thereby slightly suppressing ${\cal BR}(B \rightarrow Hb)$ and ${\cal BR}(B \rightarrow Ab)$.

To further explore this BSM scenario, we propose three BPs as detailed in Tab.~\ref{Bp4}. These BPs encapsulate the conditions discussed earlier.

\subsection{2HDM with $(XTB)$ triplet}
\begin{figure}[H]
	\centering
	\includegraphics[width=0.45\textwidth,height=0.45\textwidth]{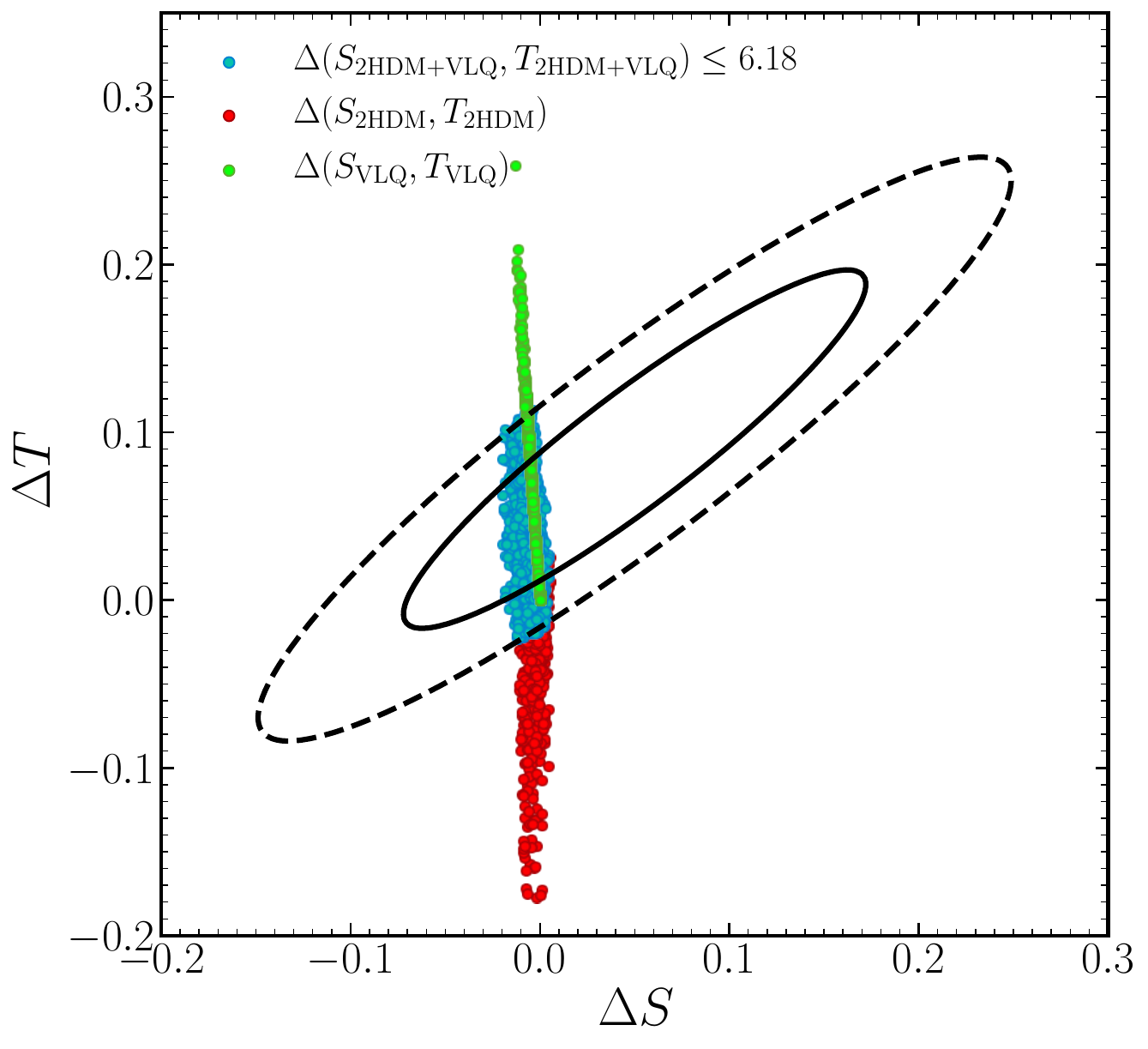}
	\caption{ Scatter plots for randomly generated points superimposed onto the 
		fit limits in the $ (\Delta S, \Delta T)$ plane from EWPO data at 95\% CL with a correlation of 92\%. Here,  we illustrate the 2HDM and VLQ contributions separately and also the total one. Further, all constraints have been taken into account.}
	\label{fig11}
\end{figure}
We now discuss the $(XTB)$ triplet scenario. Before presenting our numerical results, we will first outline our parametrization. This model is defined by the new bottom mass and one mixing angle, $\theta_L^t$, with the remaining parameters being computable. Specifically, $\theta_R^t$ is derived from Eq. (\ref{ec:rel-angle1}) while $m_X$ is given 
by \cite{Aguilar-Saavedra:2013qpa}:

\begin{eqnarray}
m_X^2=m_T^2\cos\theta^u_L+m_t^2 \sin\theta^u_L = m_B ^2 \cos^2\theta_L^b+m_b^2 \sin^2\theta_L^b.
\end{eqnarray}

Using this relationship between $m_T$ and $m_X$, along with the mixing relations in Eq.~(\ref{TBY-mix}), we can derive the mass of the new bottom quark as follows:
\begin{eqnarray}
m_B^2= \frac{1}{2}\sin^2(2\theta^u_L)(m_T^2-m_t^2)^2/(m_X^2-m_b^2)+m_X^2.
\end{eqnarray}
The down-type quark mixing is then given by.
\begin{eqnarray}
\sin(2\theta_L^d) = 
\sqrt{2}\frac{m_T^2-m_t^2}{m_B^2-m_b^2}\sin (2\theta^u_L).
\end{eqnarray}

As usual, we perform a systematic scan over the parameters of both the 2HDM and VLQ sectors, as detailed in Tab.\ref{table1}. In Fig.\ref{fig11}, it is evident that while the combined contribution of 2HDM and VLQ falls within the 95\% CL for $\Delta \chi^2$ of $\Delta S$ and $\Delta T$, the individual contributions from the 2HDM and VLQ sectors mostly lie outside the $2\sigma$ level.

\begin{figure}[H]
	\centering
	\includegraphics[width=0.85\textwidth,height=0.4\textwidth]{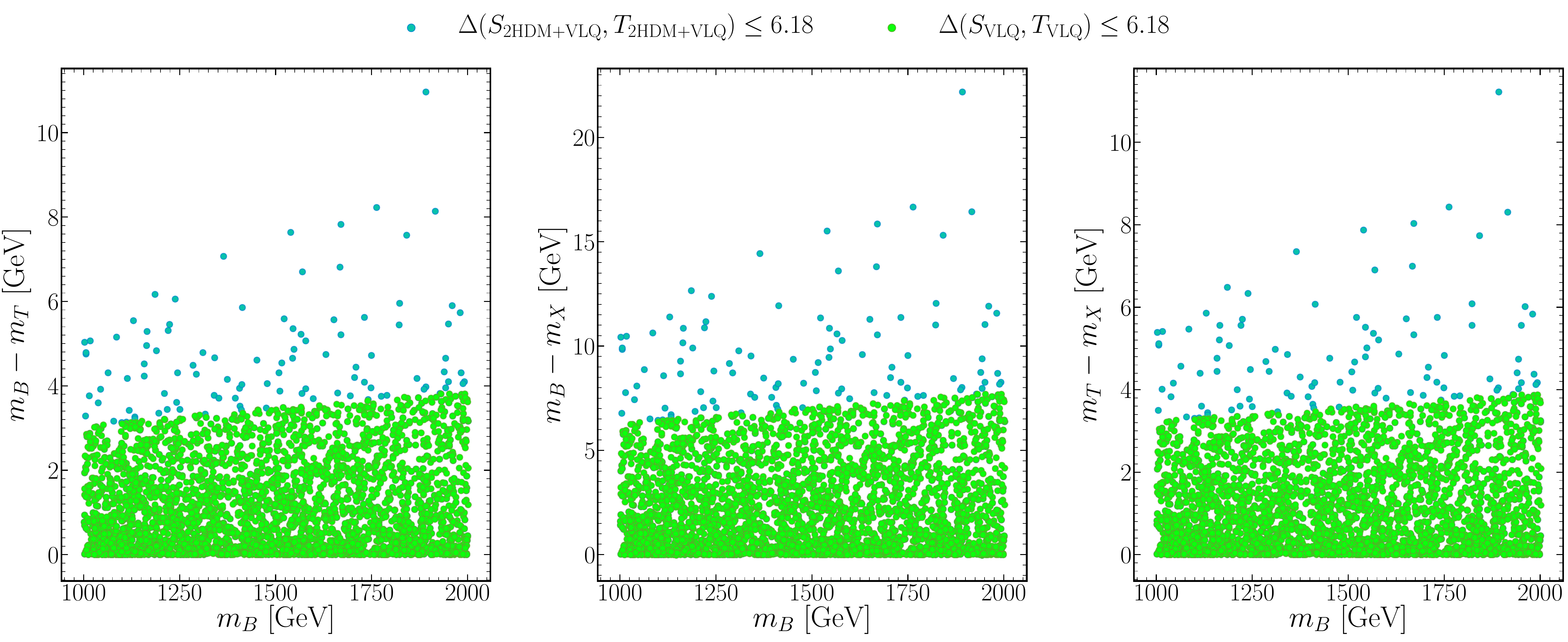}
	\caption{Scatter plots of randomly generated points mapped onto the $(m_{B},\delta)$ plane, where $\delta$ is the mass difference between $B$ and $T$, $B$ and $X$, and $T$ and $X$.  Here,  we illustrate the VLQ and 2HDM+VLQ contributions separately. Further, all constraints have been taken into account.}
	\label{fig12}
\end{figure}

Moreover, the mass splitting between $m_B$, $m_T$, and $m_X$ is tightly constrained by the EWPOs, as shown in Fig.~\ref{fig12}, where the mass differences are plotted against $m_B$. It is noticeable that the splitting is relatively small in the SM with a VLQ structure, typically a few GeV. However, in the full 2HDM with VLQ scenario, the splitting becomes significantly larger, reaching up to 11 or 22 GeV, suggesting that inter VLQ decays are likely to play a minor role in this BSM framework.

\begin{figure}[H]
	\centering
	\includegraphics[width=0.85\textwidth,height=0.4\textwidth]{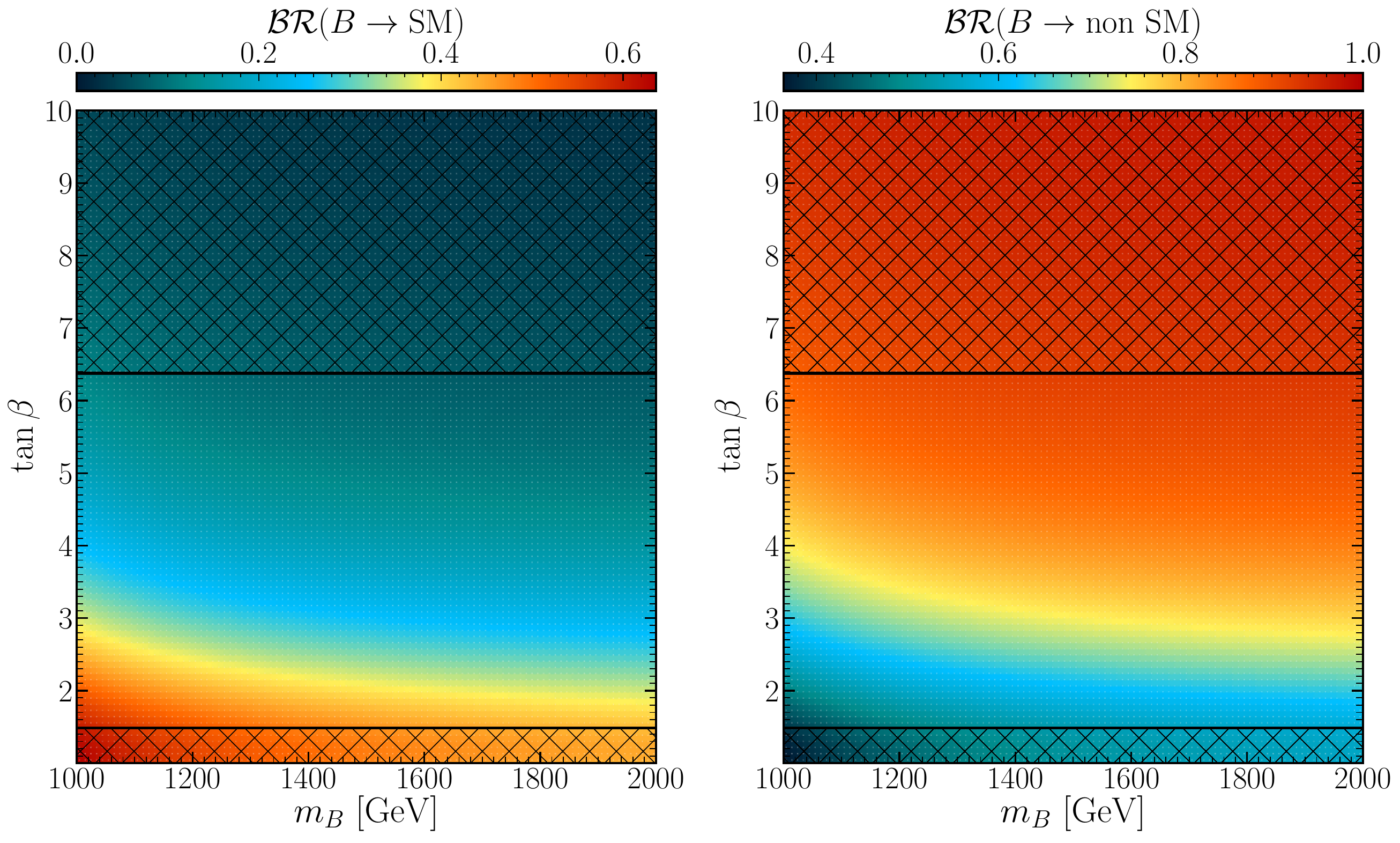} 
	\caption{The ${\cal BR}(B\to$ SM) (left) and ${\cal BR}(B\to$ non SM)  (right) mapped onto the $(m_B, \tan\beta)$ plane, with $\sin\theta_L^u=0.0093$ (the 2HDM parameters are the same as in Fig. \ref{fig3}). Here, the shaded areas are excluded by \texttt{HiggsBounds}, and all other constraints  ($S$, $T$, \texttt{HiggsSignals} and theoretical ones) are also checked.}	
	\label{fig13}
\end{figure}

In the context of the 2HDM with an $(XTB)$ triplet, the VLB can decay into SM channels (${Zb, hb}$ and ${W^-t}$) as well as additional channels involving extra Higgs states of the 2HDM (${Hb, Ab, H^-t}$). The relative importance of these decay channels is shown in Fig.~\ref{fig13}, indicating the dominance of non-SM channels, particularly at medium $\tan\beta$ values where they can reach up to 100\%. Nevertheless, SM channels can still contribute notably, achieving cumulative branching ratios of up to 60\% within the allowed parameter space.

We now turn to a detailed discussion of the individual $B$ decay channels. Note that due to the small mass difference between $B$ and $Y$, the decay $B \rightarrow W^+ Y$ is closed for a real $W^+$ and is highly suppressed for an off-shell one, thus it is not considered here.

\begin{figure}[H]
	\centering
	\includegraphics[width=0.95\textwidth,height=0.4\textwidth]{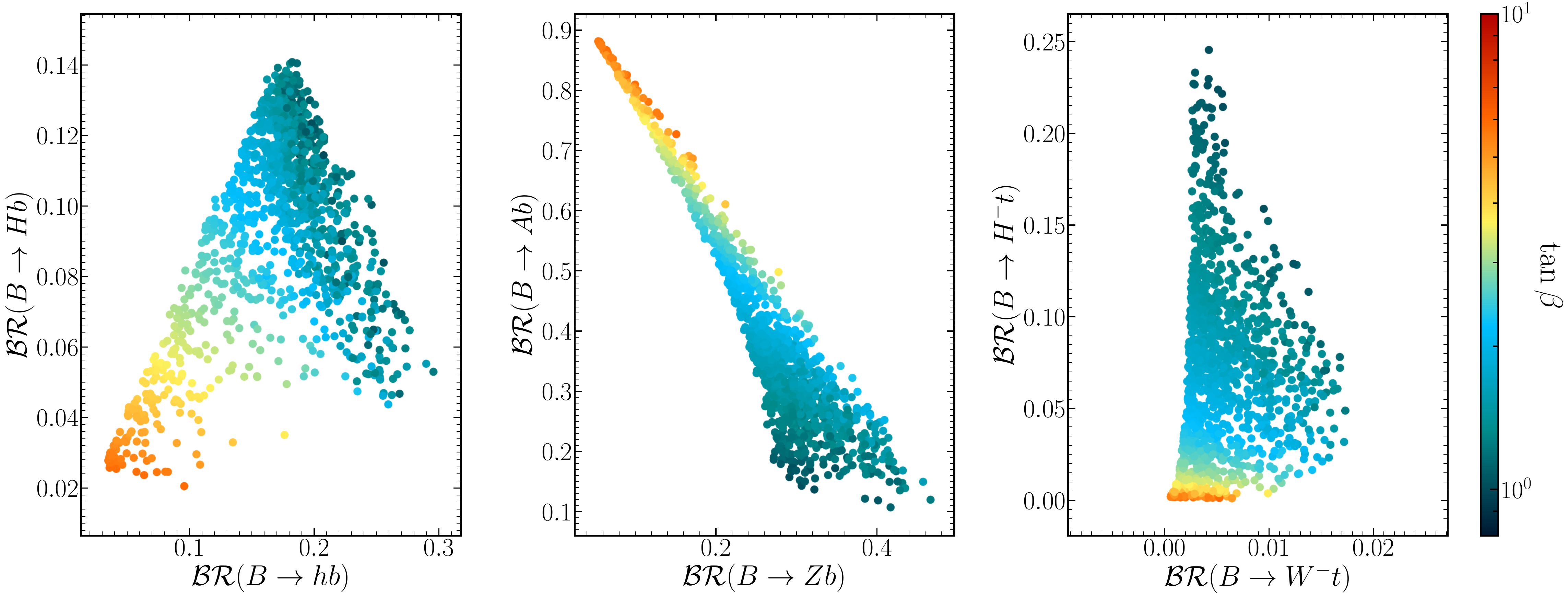}
	\caption{The correlation between ${\cal BR}(B\to hb)$ and ${\cal BR}(B\to Hb)$ (left),   
		${\cal BR}(B\to Zb)$ and ${\cal BR}(B\to Ab)$ (middle) as well as   ${\cal BR}(B\to W^-t)$ and ${\cal BR}(B\to H^-t)$ (right) with $\tan\beta$ indicated in the color gauge.}	
	\label{fig14}
\end{figure}

As previously mentioned, the inclusion of additional decay channels associated with the extra 2HDM states impacts the SM framework. While the SM predicts the decay of the new bottom quark into states such as ${Zb, hb}$ and ${W^-t}$, the full 2HDM+VLQ scenario introduces new channels (${Hb, Ab, H^-t}$) that could be significant within this multiplet structure. Specifically, the decay $B \rightarrow Ab$ could exhibit a branching ratio (${ \cal BR }$) as high as 90\% at large $\tan\beta$ values, owing to its coupling being proportional to $\tan\beta$. Conversely, the significance of other exotic decays is limited, typically reaching up to 25\%. These patterns are illustrated in Fig.~\ref{fig14}.

Lastly, we provide an overview of our proposed BPs in Tab.~\ref{Bp6}, all of which conform to the conditions discussed previously.

\begin{table}[H]
	\begin{center}
		\setlength{\tabcolsep}{30pt}
		\renewcommand{\arraystretch}{1}
		\begin{adjustbox}{max width=\textwidth}		
			\begin{tabular}{lccc}
				\toprule\toprule
				Parameters &       BP$_1$ &       BP$_2$&       BP$_3$ \\
				\toprule

				\multicolumn{4}{c}{2HDM+VLQ inputs. The masses are in GeV.} \\\toprule
				$m_h$   &   125&   125  &   125 \\
				$m_H$  &   956.83 &  792.93 &  868.07\\
				$m_A$   &813.96 &  648.12 &  627.72\\
				$m_{H\pm}$   & 882.58 &  684.27 &  703.38  \\
				$\tan\beta$ &     1.43 &    1.63 &    1.36  \\
				$m_T$      & 1030.85 & 1391.24 & 1997.91\\
				$m_B$      & 1034.61 & 1395.99 & 2006.19\\
				$m_X$ &1026.85 & 1386.32 & 1989.46\\
				$\sin(\theta^d)_L$    & 0.089 &    0.085 &    0.092\\
				$\sin(\theta^d)_R$    &  0.122 &    0.118 &    0.129\\
				\toprule
				\multicolumn{4}{c}{$\mathcal{BR}(T\to {XY})$ in \%} \\\toprule
				${\cal BR}(B\to W^+t)$  &  0.002 &  0.001 &  0.000 \\
				${\cal BR}(B\to Zb)$  &    43.32 & 25.15 &  9.74\\
				${\cal BR}(B\to hb)$  &   42.06 & 24.79 &  9.65  \\
				${\cal BR}(B\to Hb)$  &   2.69 & 22.60 & 38.38  \\
				${\cal BR}(B\to Ab)$  &   11.93 & 27.46 & 42.24\\
				${\cal BR}(B\to H^+t)$  &  0.00 &  0.00 &  0.00 \\
				\toprule
				\multicolumn{4}{c}{Total decay width in GeV.} \\\toprule
				$\Gamma(B)$ &  14.50 &   41.71 &   63.88  \\
				\toprule
				\multicolumn{4}{c}{Observables} \\\toprule
				$T_{\mathrm{2HDM}}$  &   -0.0906 &   -0.0697 &   -0.2199 \\
				$T_{\mathrm{VLQ}}$  & 0.1398 &    0.1640 &    0.2804\\
				$S_{\mathrm{2HDM}}$ & 0.0002 &    0.0027 &    0.0035 \\
				$S_{\mathrm{VLQ}}$ &    -0.0089 &   -0.0096 &   -0.0136  \\	
				
				$\Delta\chi^2(S_{\mathrm{2HDM+VLQ}},T_{\mathrm{2HDM+VLQ}})$ &    0.59 &    3.83 &    1.04\\\toprule
				$\chi^2{(h_{125})}\equiv \chi^2_{\texttt{HiggsSignals}}$ &  158.52 &  158.48 &  158.60 \\	
				
				\bottomrule\bottomrule
				
			\end{tabular}
		\end{adjustbox}
	\end{center}
	\caption{The full description of our BPs for the $(XTB)$ triplet case.}\label{Bp6}
\end{table}

\subsection{2HDM with $(TBY)$ triplet}
\begin{figure}[H]
	\centering
	\includegraphics[width=0.45\textwidth,height=0.45\textwidth]{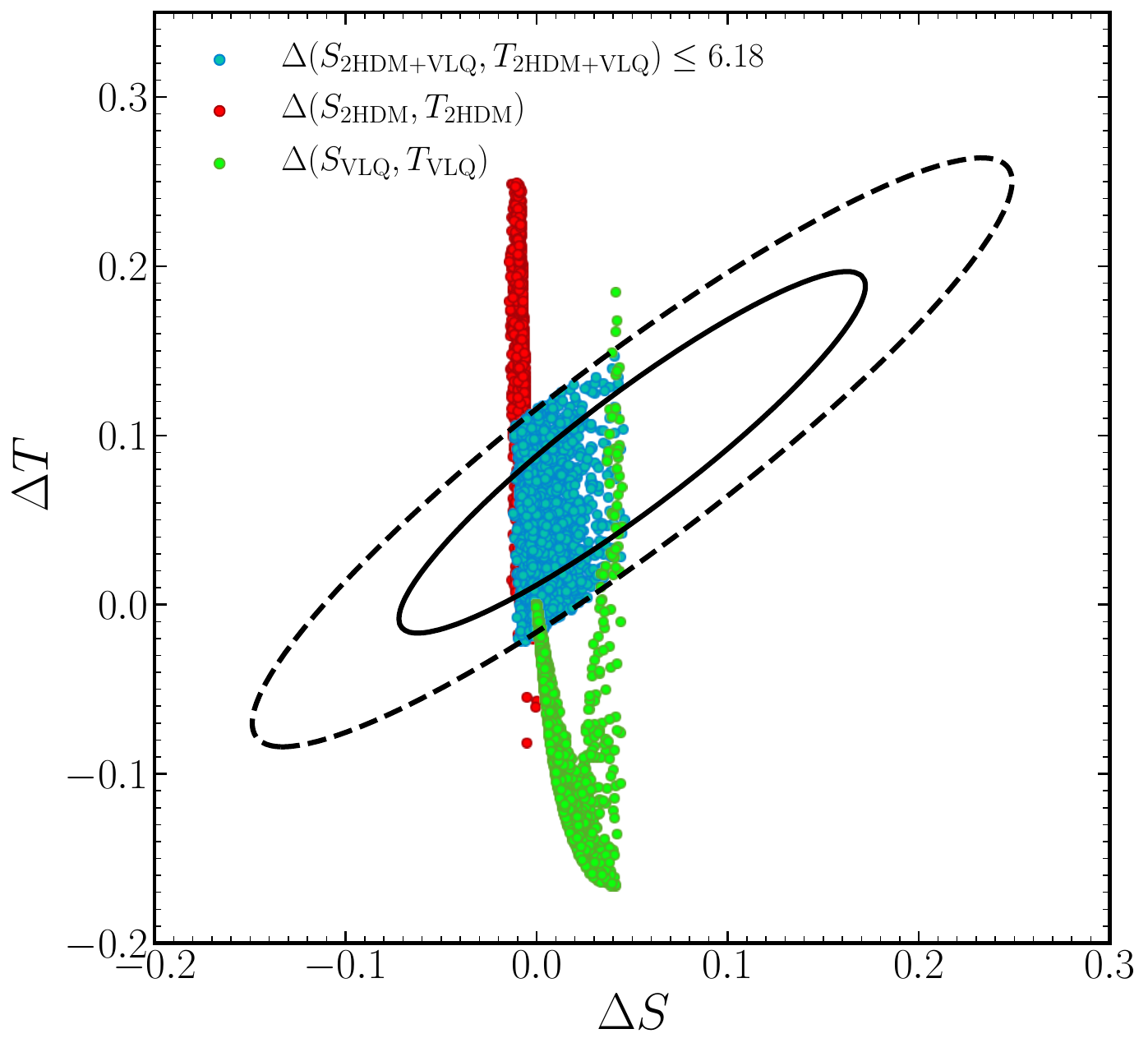}
	\caption{ Scatter plots for randomly generated points superimposed onto the 
		fit limits in the $ (\Delta S, \Delta T)$ plane from EWPO data at 95\% CL with a correlation of 92\%. Here,  we illustrate the 2HDM and VLQ contributions separately and also the total one. Further, all constraints have been taken into account.}
	\label{fig15}
\end{figure}

We conclude with the discussion of the $(TBY)$ triplet case. In the 2HDM with this VLQ representation, the setup is quite similar to the $(T)$ singlet and $(TB)$ doublet cases. Before presenting our numerical results, we first outline our parametrization. This model is defined by specifying the new top mass and one mixing angle, $\theta_L^t$, with the other parameters being derived. Specifically, $\theta_R^t$ is calculated from Eq.  (\ref{ec:rel-angle1}) while $m_Y$ is given 
by \cite{Aguilar-Saavedra:2013qpa}:
\begin{eqnarray}
m_Y^2 &=& m_T^2 \cos^2\theta_L^t+m_t^2 \sin^2\theta_L^t  = 
m_B ^2 \cos^2\theta_L^b+m_b^2 \sin^2\theta_L^b.
\end{eqnarray}
Using this relationship between $m_T$ and $m_Y$, and the mixing relations in Eq.~(\ref{TBY-mix}), the mass of the new bottom quark is derived as follows:
\begin{eqnarray}
m_B^2 &=& \frac{1}{8} \sin^2 2 \theta_L^t \frac{(m_T^2 -m_t^2)^2}{m_Y^2-m_b^2}+m_Y^2.
\end{eqnarray}
With these relations, the down-type quark mixing angles $\theta_{L,R}^d$ can be determined using  Eqs.~(\ref{ec:rel-angle1})--(\ref{TBY-mix}).
Based on the scan ranges listed in Tab.\ref{table1}, Fig.~\ref{fig15} illustrates the allowed 95\% CL regions from the $S$ and $T$ parameter constraints. It shows that while individual contributions from the 2HDM and VLQ sectors might fall outside the allowed ranges, their combined effect can produce viable solutions due to cancellations.
As mentioned earlier, the addition of VLQs and extra Higgs states can significantly alter the phenomenology of the $S$ and $T$ parameters compared to the SM scenario.
\begin{figure}[H]
	\centering
	\includegraphics[width=0.85\textwidth,height=0.4\textwidth]{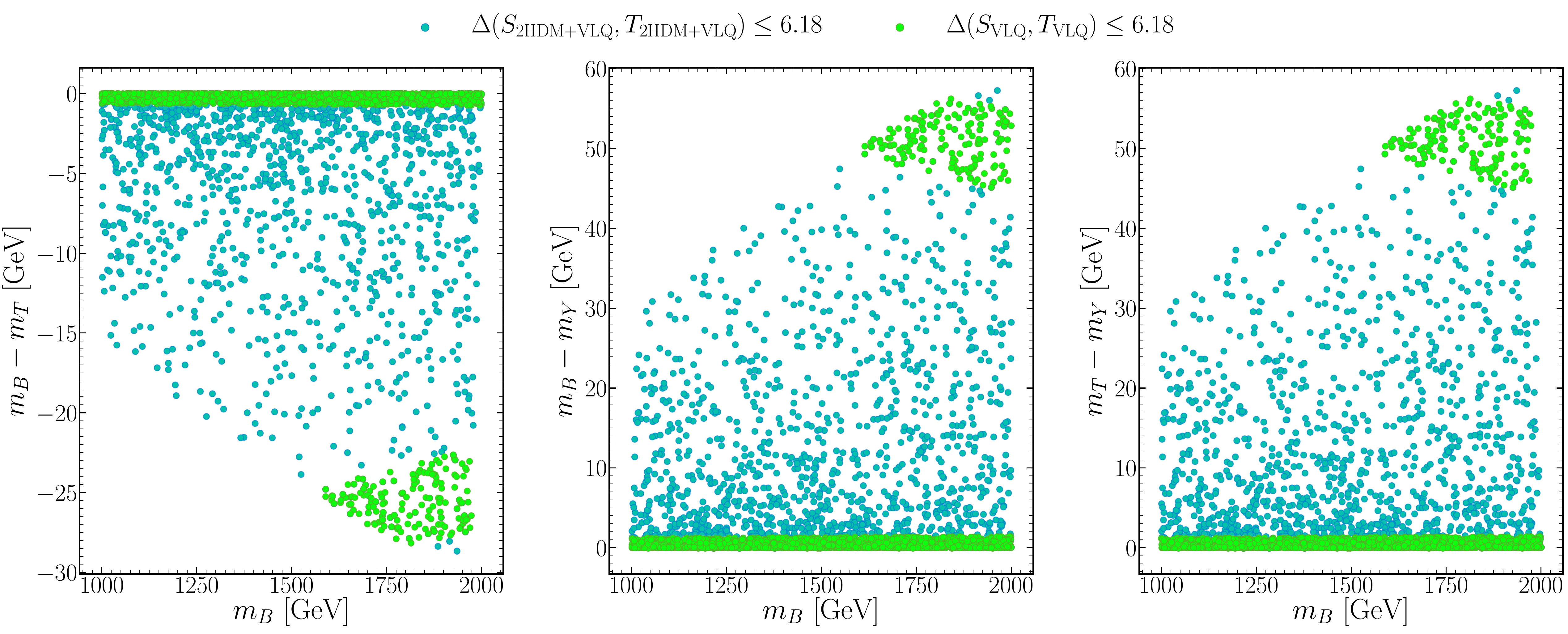}
	\caption{Scatter plots of randomly generated points mapped onto the $(m_{T/B},\delta)$ plane, where $\delta$ is the mass difference between $T$ and $B$, $T$ and $Y$, and $B$ and $Y$.  Here,  we illustrate the VLQ and 2HDM+VLQ contributions separately. Further, all constraints have been taken into account.}
	\label{fig16}
\end{figure}
In Fig.~\ref{fig16}, we analyze the mass splitting $\delta$ between the masses of $T$ and $B$ (left), $T$ and $Y$ (middle), and $B$ and $Y$ (right) within the SM extended by this VLQ representation. The splitting is generally small, on the order of a fraction of a GeV at low $m_T$. For higher $m_T$ values, there is a narrow parameter space where these splittings could be around 1 GeV, which is still minimal.

In contrast, in the 2HDM+VLQ scenario, the dynamics change considerably. The splitting between $B$ and $T$ can reach $-15$ GeV for low $m_B$ (around 1000 GeV), while the splittings between $B$ and $Y$ and $T$ and $Y$ can extend up to 30 GeV in the same mass range. For larger $m_B$, the mass splittings in both the individual VLQ contributions and the combined 2HDM+VLQ scenario can reach up to $-28$ GeV between $B$ and $T$, and up to $58$ GeV between $B$ and $Y$ and $T$ and $Y$.

Once again, significant cancellations between additional Higgs and VLQ states, occurring naturally in loop contributions due to their different spin statistics, expand the viable parameter space. This expanded parameter space will be further explored in accordance with the analyses presented in the preceding subsections to study 2HDM decays of VLQ states. However, the small values of $\delta=m_i-m_j$ observed in these plots highlight that VLQ decays into each other are mostly negligible.

\begin{figure}[H]
	\centering
	\includegraphics[width=0.85\textwidth,height=0.4\textwidth]{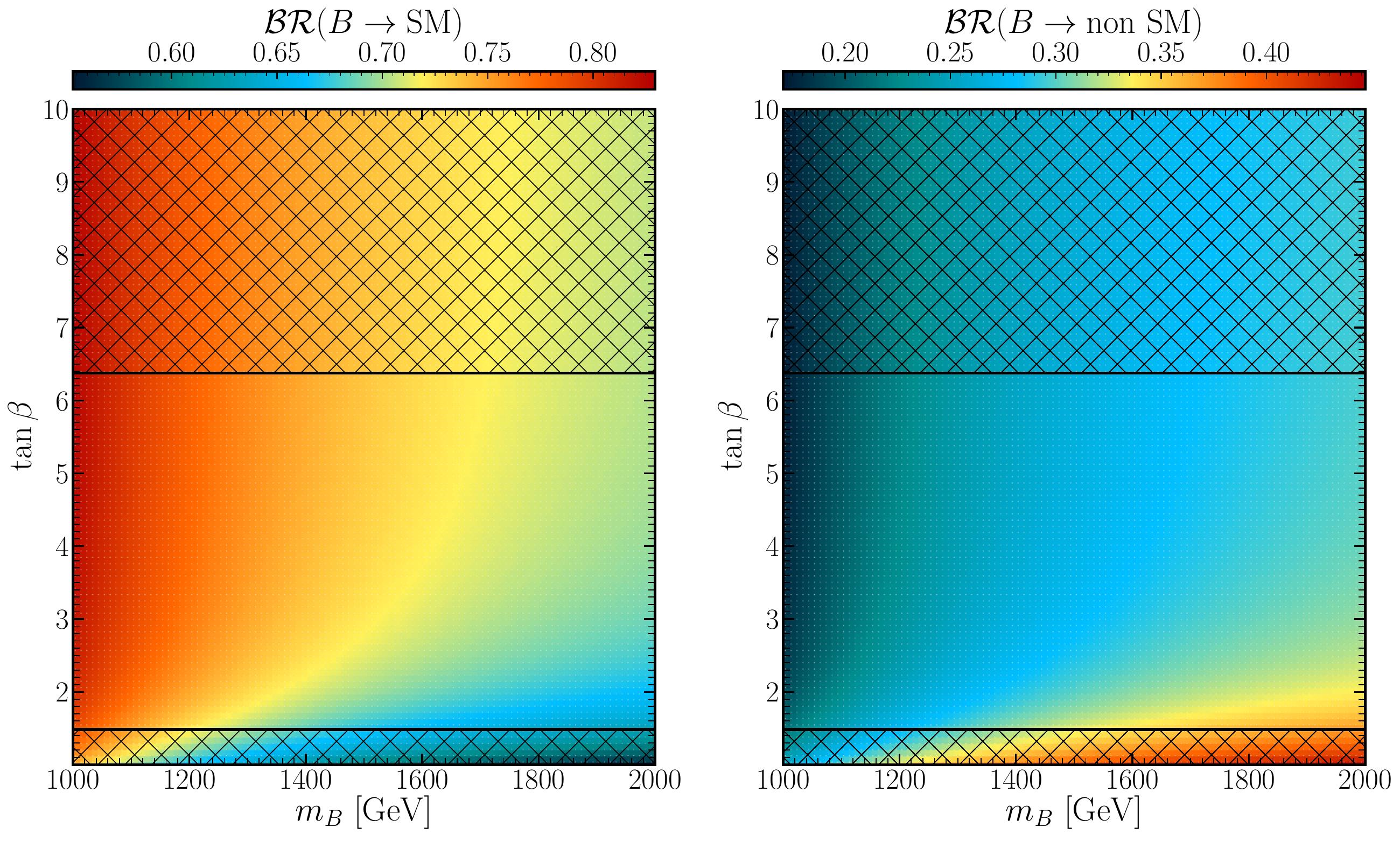} 
	\caption{The ${\cal BR}(T\to$ SM) (left) and ${\cal BR}(T\to$ non SM)  (right) mapped onto the $(m_T, \tan\beta)$ plane, with $\sin\theta_L^u=0.02$ (the 2HDM parameters are the same as in Fig. \ref{fig3}). Here, the shaded areas are excluded by \texttt{HiggsBounds}, and all other constraints  ($S$, $T$, \texttt{HiggsSignals} and theoretical ones) are also checked.}	
	\label{fig17}
\end{figure}

\begin{figure}[H]
	\centering
	\includegraphics[width=1\textwidth,height=0.4\textwidth]{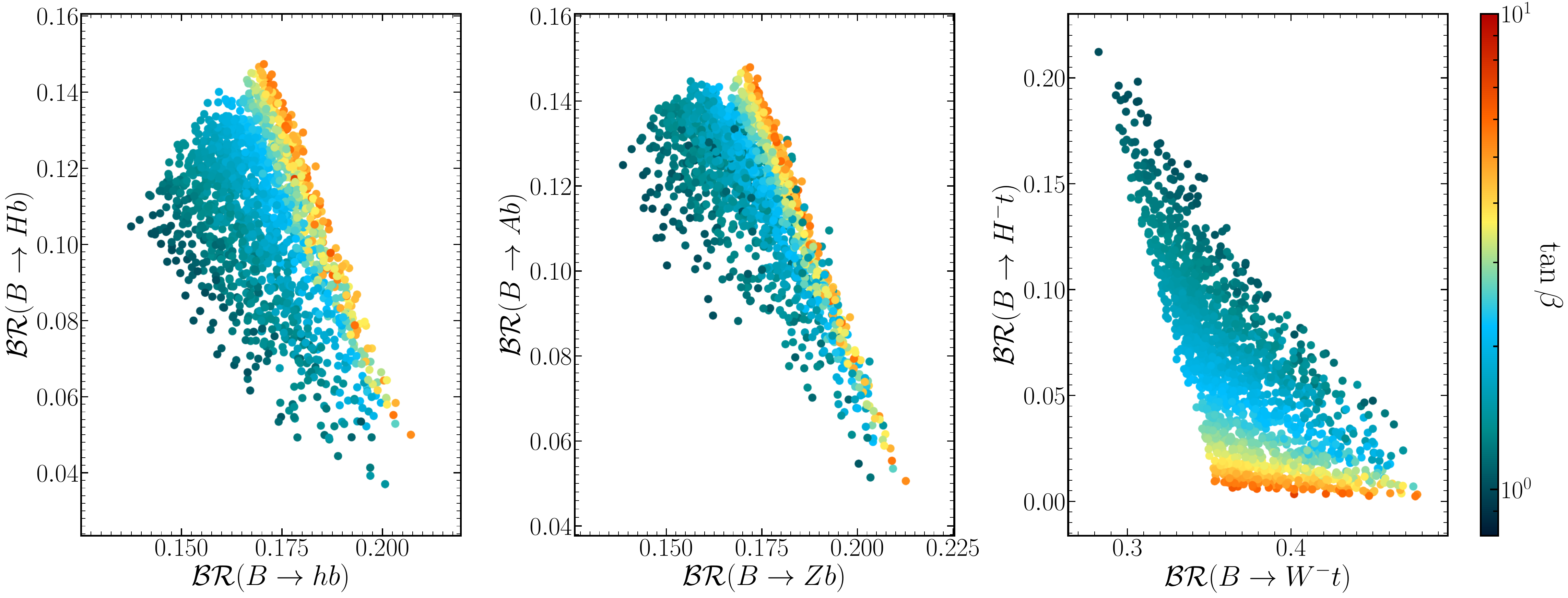}
	\caption{The correlation between  ${\cal BR}(B\to hb)$ and ${\cal BR}(B\to Hb)$ (left)  (right),   
		${\cal BR}(B\to Zb)$ and ${\cal BR}(B\to Ab)$ (middle) as well as  ${\cal BR}(B\to W^-t)$ and ${\cal BR}(B\to H^-t)$(left) with $\tan\beta$ indicated in the color gauge.}	
	\label{fig18}
\end{figure}

In the SM with a $(TBY)$ triplet, the decay patterns of $B$ states are essentially the same as in the $(TB)$ doublet case. Therefore, in Fig.~\ref{fig17}, we investigate the decays of the new bottom quark into both SM channels (${Zb, hb}$ and ${W^-t}$) as well as exotic ones (${Hb, Ab}$ and ${H^-t}$). The cumulative branching ratios (${ \cal BR }$) of non-SM decays are notably pronounced, particularly evident at small $\tan\beta$ values and large $m_B$, where they can reach up to 35\% within the allowed region. However, SM channels still dominate, accounting for approximately 80\% of cumulative ${\cal BR}$s for small $m_B$ values around 1000 GeV within the allowed range.

We now delve into the individual $B$ decay channels depicted in Fig.~\ref{fig18}. It is important to note that, due to the small splitting between $B$ and $Y$, the decay $B \rightarrow W^- Y$ is closed for a real $W^-$ and highly suppressed for an off-shell one, so we will not discuss it here. Clearly, the figure shows a strong anti-correlation between standard and exotic channels. For charged currents, the $\tan\beta$ dependence reveals that large values favor SM decays, whereas small values favor 2HDM ones. Regarding neutral currents, both large and medium $\tan\beta$ values result in SM and 2HDM decays achieving their maxima, around 21\% for the SM and 15\% for the 2HDM.

\begin{figure}[H]
	\centering
	\includegraphics[width=0.45\textwidth,height=0.45\textwidth]{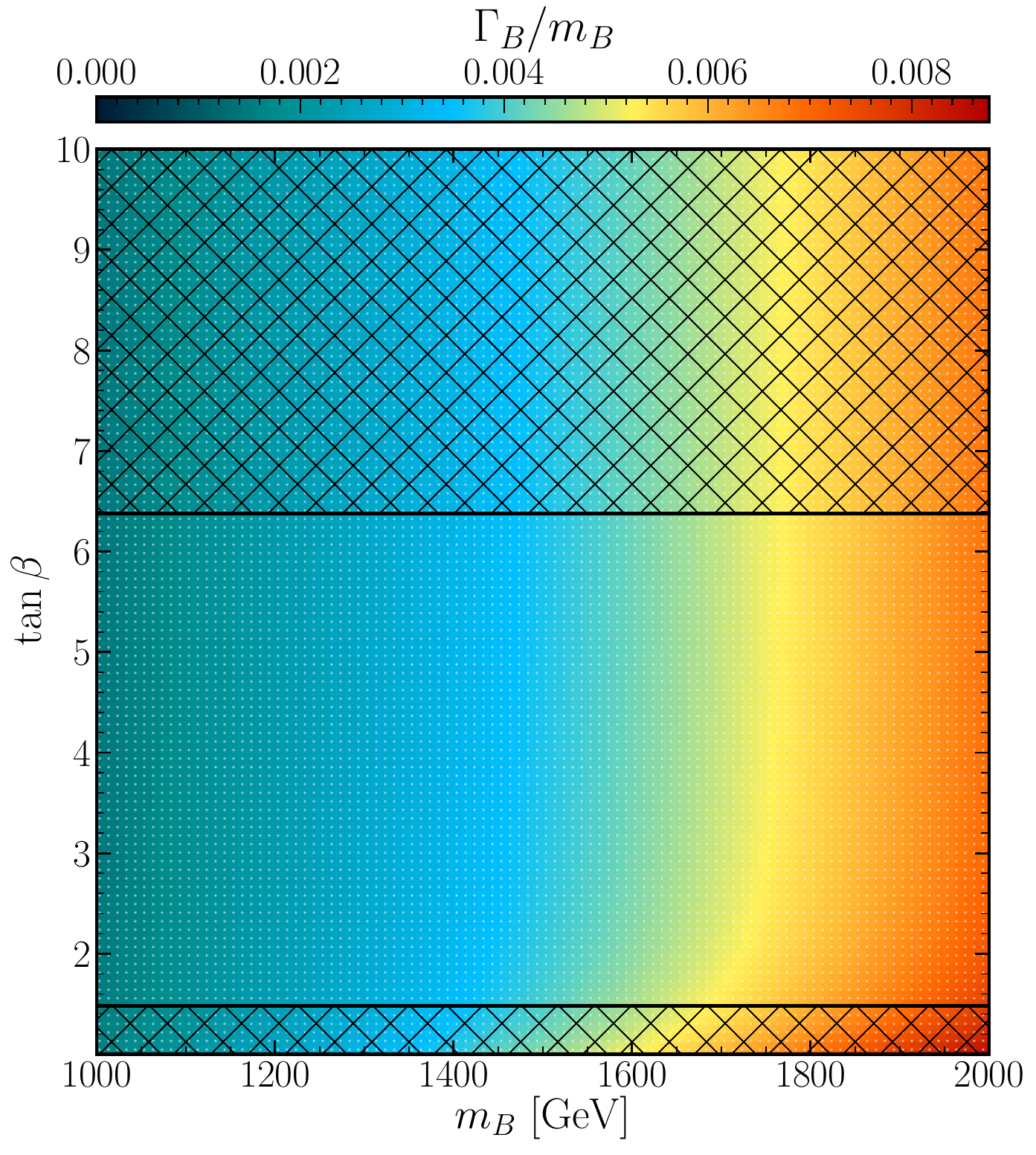} 
	\caption{The ratio $\Gamma(B)/m_B$ mapped over the $(m_T,\tan\beta)$ plane.  with $\sin\theta_L^u=0.02$ (the 2HDM parameters are the same as in Fig. \ref{fig3}). Here, the shaded areas are excluded by \texttt{HiggsBounds}, and all other constraints  ($S$, $T$, \texttt{HiggsSignals} and theoretical ones) are also checked.}	
	\label{fig19}
\end{figure}

Proceeding to our BPs, we first analyze the ratio $\Gamma(B)/m_B$ illustrated in Fig.~\ref{fig19}. This visualization underscores the consistently narrow nature of the $B$ state within our BSM framework. Notably, $\Gamma(B)/m_B$ reaches its peak of 0.008, primarily observed for high values of $m_B$.

Finally, in Tab.~\ref{Bp7}, we present our BPs explicitly, which include both light (BP$_1$) and heavy (BP$_2$) $B$ states. This diverse selection allows for a comprehensive examination of exotic $B$ decays in phenomenological analyses.	

\begin{table}[h!]
	\begin{center}
		\setlength{\tabcolsep}{45pt}
		\renewcommand{\arraystretch}{0.8}
		\begin{adjustbox}{max width=\textwidth}		
			\begin{tabular}{lcc}
				\toprule\toprule
				Parameters &       BP$_1$ &       BP$_2$ \\
				\toprule

				\multicolumn{3}{c}{2HDM+VLQ inputs. The masses are in GeV.} \\\toprule
				$m_h$   &   125&   125  \\
				$m_H$  &   623.83 &  702.87\\
				$m_A$   &624.27 &  631.34 \\
				$m_{H\pm}$   &745.51 &  766.97  \\
				$\tan\beta$ &    1.08 &    1.03 \\
				$m_T$      & 1129.42 & 1984.92\\
				$m_B$      & 1120.77 & 1968.01\\
				$m_Y$      & 1112.46 & 1951.23\\				
				$\sin(\theta^u)_L$    & -0.175 &    0.184 \\
				$\sin(\theta^d)_L$    &-0.122 &    0.130\\
				
				\toprule
				\multicolumn{3}{c}{$\mathcal{BR}(T\to {XY})$ in \%} \\\toprule
				${\cal BR}(B\to W^+t)$  &  39.04 & 29.29  \\
				${\cal BR}(B\to W^+T)$  &   - &   - \\
				${\cal BR}(B\to Zb)$  &   17.76 & 14.34 \\
				${\cal BR}(B\to hb)$  &   17.33 & 14.22  \\
				${\cal BR}(B\to Hb)$  &   8.46 & 10.91    \\
				${\cal BR}(B\to Ab)$  &  8.45 & 11.54 \\
				${\cal BR}(B\to H^+t)$ &  8.96 & 19.71 \\
				${\cal BR}(B\to H^+T)$ &   - &  -   \\				
				\toprule
				\multicolumn{3}{c}{Total decay width in GeV.} \\\toprule
				$\Gamma(B)$ &  18.92 &  145.66  \\
				\toprule
				\multicolumn{3}{c}{Observables} \\\toprule
				
				$T_{\mathrm{2HDM}}$  &  0.2624 &    0.1546\\
				$T_{\mathrm{VLQ}}$  &  -0.1415 &   -0.0843\\
				$S_{\mathrm{2HDM}}$ &  -0.0092 &   -0.0072\\
				$S_{\mathrm{VLQ}}$ &   0.0212 &    0.0244  \\	
				$\Delta\chi^2(S_{\mathrm{2HDM+VLQ}},T_{\mathrm{2HDM+VLQ}})$ &     5.25 &    0.23\\\toprule
				$\chi^2{(h_{125})}\equiv \chi^2_{\texttt{HiggsSignals}}$ & 158.45 &  158.13 \\	
				
				\bottomrule\bottomrule
				
			\end{tabular}
		\end{adjustbox}
	\end{center}
	\caption{The full description of our BPs for the $(TBY)$ triplet case.}\label{Bp7}
\end{table}
\section{Conclusions}

The third-generation quarks of the SM, top and bottom, are often associated with SM extensions involving VLQs with the same EM charge, which can mix and interact with them, thereby affecting their phenomenology. This approach has typically been explored within the context of a SM-like Higgs sector, but the relevance of these new states could be even greater in BSM scenarios featuring an extended Higgs sector. In such scenarios, VLQs and additional Higgs states can strongly interact with each other. Furthermore, VLQs with exotic charges can be part of the same representation, adding complexity to the model's phenomenology. 

In this paper, we considered a Type-II 2HDM supplemented by a VLB companion and other new fermionic states ($T$, $X$, and $Y$), arranged in singlet, doublet, or triplet representations. After constraining the parameter space of this model for all VLQ multiplet cases against the latest theoretical and experimental constraints, we analyzed the ‘standard’ decays $B \to W^-t, Zb$, and $hb$ (where $h$ is the discovered SM-like Higgs state) against the ‘exotic’ decays $B \to H^-t, Ab$, and $Hb$ (where $H^-$, $A$, and $H$ are the heavy Higgs states of the 2HDM). It is important to note that $T$ states were not considered among the decay products of the $B$ states due to the strong mass degeneracy, which also involves the exotic VLQ states $X$ and $Y$.

Our conclusions are based on a comprehensive analysis conducted at the branching ratio (${ \cal BR }$) level, without Monte Carlo simulations. Our exploration of parameter space revealed promising scenarios where $B$, $H^-$, $A$, and $H$ particles can be light enough to be probed in direct searches at the LHC, following $B\overline B$ pair production via QCD. Additionally, our investigation into the exotic decay modes of $B$ and the assessment of their relative rates provided valuable insights into the underlying structure of the VLQ multiplet. This enhanced understanding significantly contributes to characterizing the BSM scenario, including confirming its Type-II nature and measuring key parameters within our 2HDM+VLQ model.

{The decays of the heavy Higgs states $H$, $A$, and $H^-$ offer a distinct final state at the LHC. Specifically, $H$ and $A$ predominantly decay into $t\bar{t}$ pairs, while $H^-$ primarily decays into $\bar t{b}$, resulting in multiple $W$ bosons and $b$-jets in the final state. In the case of $B\overline{B}$ pair production, this leads to a final state with $4W + 6b$, providing a rich bottom-quark signature. Additionally, if two same-sign $W$ bosons decay leptonically, the $4W+6b$ final state could produce same-sign leptons.}

We advocate for further exploration through Monte Carlo simulations as the next step in our research program. This initiative aims to rigorously test the feasibility of both an extended Higgs sector and an augmented fermionic spectrum being simultaneously accessible at the LHC through non-SM signatures of VLQs produced in pairs in QCD collisions. To this end, we have proposed several BPs suitable for experimental investigation during Run 3 of the LHC and/or its high-luminosity phase.

	\section*{Acknowledgments}
	The work of AB, RB and MB is supported by the Moroccan Ministry of Higher Education and Scientific Research MESRSFC and CNRST Project PPR/2015/6. SM is supported in part through the NExT Institute,  STFC Consolidated Grant ST/L000296/1 and  Knut and Alice Wallenberg foundation under the grant KAW 2017.0100 (SHIFT).

	\section*{Appendix}
	\renewcommand{\thetable}{\Roman{table}} 
	\subsection{Lagrangian in the mass basis}
	
	After EWSB, we are left with five Higgs bosons that are two-CP even $h$ and $H$, one CP-odd $A$, and then a pair of charged Higgs $H^\pm$. We now collect the Lagrangian on the mass basis in the general 2HDM Type-II. 
	\relscale{0.95}
\subsection{Light-light interactions}
\begin{eqnarray}
\mathcal{L}_W &=& - \frac{g}{\sqrt{2}} \overline{t} \gamma^\mu (V^L_{tb}P_L + V^R_{tb} P_R)b W^+_\mu + H.c., \nonumber\\
\mathcal{L}_Z &=& - \frac{g}{2c_W}	 \overline{t} \gamma^\mu ( X^L_{tt}P_L + X^R_{tt}P_R - 2 Q_t s_W^2 ) t Z_\mu \nonumber \\ 
&& - \frac{g}{2c_W}	 \overline{b} \gamma^\mu (- X^L_{bb}P_L - X^R_{bb}P_R - 2 Q_b s_W^2 ) b Z_\mu   + H.c., \nonumber\\
\mathcal{L}_{h^0} &=& - \frac{g m_t}{2 M_W} Y_{tt}^h \overline{t} t h^0 - \frac{g m_b}{2 M_W} Y_{bb}^h \overline{b} b h^0   + H.c., \nonumber\\
\mathcal{L}_{H^0} &=& -  \frac{g m_t}{2 M_W} Y_{tt}^H \overline{t} t H^0 - \frac{g m_b}{2 M_W} Y_{bb}^H \overline{b} b H^0  + H.c.,\nonumber\\
\mathcal{L}_{A} &=& - i \frac{g m_t}{2 M_W} Y_{tt}^A \overline{t}  \gamma_5 t A + i\frac{g m_b}{2 M_W} Y_{bb}^A \overline{b} \gamma_5 b A  + H.c., \nonumber \\
\mathcal{L}_{H^+} &=& -\frac{gm_t}{ \sqrt{2} M_W} \overline{t} (\cot\beta Z^L_{tb} P_L + \tan\beta Z^R_{tb}P_R)b H^+ + H.c.
\end{eqnarray}
\begin{eqnarray}
\begin{array}{c|ccc}
& V^L_{tb} && V^R_{tb} \\ \hline
(B)	& c_L && 0 \\ 
(TB)&  c_L^u c_L^d + s_L^u s_L^d e^{i(\phi_u - \phi_d)}   && s_R^u s_R^d e^{i(\phi_u - \phi_d)} \\
(BY)& c_L &&0 \\ 
(XTB)&  c_L^u c_L^d + \sqrt{2} s_L^u s_L^d   && \sqrt{2} s_R^u s_R^d    \\ 
(TBY)&  c_L^u c_L^d + \sqrt{2} s_L^u s_L^d    && \sqrt{2} s_R^u s_R^d  
\end{array} 
\nonumber
\end{eqnarray}	

\captionof{table}{Light-light couplings to the $W$ boson.}	

\begin{eqnarray}
\begin{array}{c|cccccc}
& \quad	X_{tt}^L &&\quad	X_{tt}^R&\quad	X_{bb}^L  && \quad X_{bb}^R  \\ \hline
(B)	& 1 && 0  &c_L^2  && 0\\
(TB) & 1 && (s_R^u)^2  &1 && (s_R^d)^2 \\
(BY)	& 1 && 0  & c_L^2 - s_L^2  && -s_R^2\\
(XTB)	& (c_L^u)^2 && 0  & 1 + (s_L^d)^2  && 2 (s_R^d)^2\\
(TBY)	& 1 + (s_L^u)^2   && 2 (s_R^u)^2  & (c_L^d)^2 && 0
\end{array} \nonumber
\end{eqnarray}\captionof{table}{Light-light couplings to the $Z$ boson.}

\begin{eqnarray}
\begin{array}{c|ccc }
&  	Y_{tt}^h & 	Y_{tt}^H& 	Y_{tt}^A    \\ \hline
(B)	& s_{\beta\alpha}  +    c_{\beta\alpha} \cot\beta & c_{\beta\alpha}  -    s_{\beta\alpha} \cot\beta & -\cot\beta  \\
(TB) & (s_{\beta\alpha}  +    c_{\beta\alpha} \cot\beta) (c_R^u)^2 & (c_{\beta\alpha}  -    s_{\beta\alpha} \cot\beta) (c_R^u)^2  &-\cot\beta (c_R^u)^2\\
(BY)	& s_{\beta\alpha}  +    c_{\beta\alpha} \cot\beta & c_{\beta\alpha}  -    s_{\beta\alpha} \cot\beta & -\cot\beta  \\
(XTB)	& (s_{\beta\alpha}  +    c_{\beta\alpha} \cot\beta)(c_L^u)^2    &  (c_{\beta\alpha}  -    s_{\beta\alpha} \cot\beta) (c_L^u)^2    & -\cot\beta (c_L^u)^2   \\
(TBY)	& (s_{\beta\alpha}  +    c_{\beta\alpha} \cot\beta)(c_L^u)^2    &  (c_{\beta\alpha}  -    s_{\beta\alpha} \cot\beta) (c_L^u)^2    & -\cot\beta (c_L^u)^2   
\end{array} \nonumber
\end{eqnarray}\captionof{table}{Light-light top quark couplings to the triplets Higgs \{$h,H,A$\}.}

\begin{eqnarray}
\begin{array}{c|ccc }
&  	Y_{bb}^h & 	Y_{bb}^H& 	Y_{bb}^A    \\ \hline
(B)	& (s_{\beta\alpha}  -    c_{\beta\alpha} \tan\beta) c_L^2 &  (c_{\beta\alpha}  +   s_{\beta\alpha} \tan\beta) c_L^2   & \tan\beta c_L^2 \\
(TB) & (s_{\beta\alpha}  -    c_{\beta\alpha} \tan\beta) (c_R^d)^2 & (c_{\beta\alpha}  +   s_{\beta\alpha} \tan\beta) (c_R^d)^2  & \tan\beta (c_R^d)^2\\
(BY)	& (s_{\beta\alpha}  -    c_{\beta\alpha} \tan\beta) c_R^2 &  (c_{\beta\alpha}  +   s_{\beta\alpha} \tan\beta) c_R^2   & \tan\beta c_R^2 \\
(XTB)	& (s_{\beta\alpha}  -    c_{\beta\alpha} \tan\beta) (c_L^d)^2 &  (c_{\beta\alpha}  +   s_{\beta\alpha} \tan\beta) (c_L^d)^2  & \tan\beta (c_L^d)^2 \\
(TBY)	& (s_{\beta\alpha}  -    c_{\beta\alpha} \tan\beta) (c_L^d)^2 &  (c_{\beta\alpha}  +   s_{\beta\alpha} \tan\beta) (c_L^d)^2  & \tan\beta (c_L^d)^2 \\
\end{array} \nonumber
\end{eqnarray}\captionof{table}{Light-light bottom quark couplings to the triplets Higgs \{$h,H,A$\}.}

\begin{eqnarray}
\begin{array}{c|ccc}
& Z^L_{tb} && Z^R_{tb} \\ \hline
(B)	& c_L  && \frac{m_t}{m_B}c_L \\ 
(TB)&  c_L^d c_L^u + \frac{s_L^d}{s_L^u } (s_L^u{}^2 - s_R^u{}^2) e^{i(\phi_u - \phi_d)}   && \frac{m_b}{m_t}\left[ c_L^u c_L^d + \frac{s_L^u}{s_L^d } (s_L^d{}^2 - s_R^d{}^2) e^{i(\phi_u - \phi_d)} \right]  \\
(BY)& c_R  &&  c_L  \\ 
(XTB)&  c_L^u   &&  c_L^d   \\ 
(TBY)&     c_L^u   &&  c_L^d 
\end{array} 
\nonumber
\end{eqnarray}	
\captionof{table}{Light-light couplings to the charged Higgs.}	

\subsection{Heavy-heavy interactions}	

\begin{eqnarray}
\mathcal{L}_W&=& - \frac{g}{\sqrt{2}} \overline{Q} \gamma^\mu (V^L_{QQ^\prime} P_L + V^R_{QQ^\prime} P_R    ) Q^\prime W^+_\mu + H.c.,\nonumber\\
\mathcal{L}_Z &=& - \frac{g}{2 c_W}  \overline{Q} \gamma^\mu (\pm X^L_{QQ}P_L\pm X^R_{QQ}P_R-2Q_Qs_W^2  )Q Z_\mu \nonumber \\\
\mathcal{L}_{h^0} &=& - \frac{g m_Q}{2 M_W} Y_{QQ}^h \overline{Q} Q h^0   + H.c., \nonumber\\
\mathcal{L}_{H^0} &=& -  \frac{g m_Q}{2 M_W} Y_{QQ}^H \overline{Q} Q H^0   + H.c.,\nonumber\\
\mathcal{L}_{A} &=& \pm i \frac{g m_Q}{2 M_W} Y_{QQ}^A \overline{Q}  \gamma_5 Q A  + H.c., \nonumber \\
\mathcal{L}_{H^+} &=& -\frac{gm_Q}{ \sqrt{2} M_W} \overline{Q} (\cot\beta Z^L_{QQ} P_L + \tan\beta Z^R_{QQ}P_R)Q H^+ + H.c.
\end{eqnarray}

\begin{eqnarray}
\begin{array}{c|cccc}
& 		V_{ {T}B}^L &\quad	V_{ {T}B}^R & 	V_{ BY}^L &	V_{BY}^R \\ \hline
(TB)	& \quad s_L^u s_L^de^{-i(\phi_u - \phi_d)} + c_L^u c_L^d   & \quad  c_R^u c_R^d &-&- \\
(BY)	 & - & \quad  - &c_L& c_R \\
(XTB)	& s_L^u s_L^d + \sqrt{2} c_L^u c_L^d & \quad \sqrt{2} c_R^u c_R^d  &- & -\\
(TBY)	 & s_L^u s_L^d + \sqrt{2} c_L^u c_L^d & \quad \sqrt{2} c_R^u c_R^d  & \sqrt{2} c_L^d & \sqrt{2} c_R^d
\end{array} \nonumber
\end{eqnarray}\captionof{table}{Heavy-heavy couplings to the $W$ boson.}	

\begin{eqnarray}
\begin{array}{c|ccccc}
& 	X_{BB}^L &	X_{BB}^R  \\ \hline

(B)	  & (s_L)^2 &&&& 0  \\

(TB)	&1  &&&& (c_R^d)^2 \\
(BY)& s_L^2 - c_L^2 &&&& - c_R^2	 \\
(XTB)& 1 + (c_L^d)^2  &&&& 2 (c_R^d)^2 	 \\
(TBY)&   (s_L^d)^2 &&&& 	0
\end{array} \nonumber
\end{eqnarray}\captionof{table}{Heavy-heavy couplings to the $Z$ boson.}

\begin{eqnarray}
\begin{array}{c|cccc }
&  	Y_{BB}^h & 	Y_{BB}^H& 	Y_{BB}^A&Y_{YY}^{\{h,H,A\}}     \\ \hline 
(B)	& (s_{\beta\alpha}  -    c_{\beta\alpha} \tan\beta) s_L^2 &  (c_{\beta\alpha}  +   s_{\beta\alpha} \tan\beta) s_L^2   & \tan\beta s_L^2 &- \\
(TB) & (s_{\beta\alpha}  -    c_{\beta\alpha} \tan\beta) (s_R^d)^2 & (c_{\beta\alpha}  +   s_{\beta\alpha} \tan\beta) (s_R^d)^2  & \tan\beta (s_R^d)^2&- \\
(BY)	& (s_{\beta\alpha}  -    c_{\beta\alpha} \tan\beta) s_R^2 &  (c_{\beta\alpha}  +   s_{\beta\alpha} \tan\beta) s_R^2   & \tan\beta s_R^2&0\\
(XTB)	& (s_{\beta\alpha}  -    c_{\beta\alpha} \tan\beta) (s_L^d)^2 &  (c_{\beta\alpha}  +   s_{\beta\alpha} \tan\beta)(s_L^d)^2   & \tan\beta  (s_L^d)^2 & - \\
(TBY)	& (s_{\beta\alpha}  -    c_{\beta\alpha} \tan\beta) (s_L^d)^2 &  (c_{\beta\alpha}  +   s_{\beta\alpha} \tan\beta)(s_L^d)^2   & \tan\beta  (s_L^d)^2 &0
\end{array} \nonumber
\end{eqnarray}\captionof{table}{Heavy-heavy Bottom VLQ couplings to the triplets Higgs \{$h,H,A$\}.}

\begin{eqnarray}
\begin{array}{c|ccccc}
&  Z^L_{TB} && Z^R_{TB}& Z^L_{BY} & Z^R_{BY} \\ \hline 
(TB)& s_L^d s_L^u e^{i(\phi_d - \phi_u)}+ \frac{ c_L^d}{c_L^u} (s_R^u{}^2    -      s_L^u{}^2 ) &&\frac{m_B}{ m_T}   \left[s_L^u s_L^d e^{i(\phi_d - \phi_u)}+  \frac{c_L^u}{ c_L^d} ( s_R^d{}^2    -      s_L^d{}^2 )  \right] &-&- \\ 
(XTB)& -   && - &-&-  \\ 
(TBY)& -   &&- &s_L e^{-i\phi}&0  
\end{array} 
\nonumber
\end{eqnarray}	
\captionof{table}{Heavy-heavy couplings to the charged Higgs.}

\subsection{Light-heavy interactions}

\begin{eqnarray}
\mathcal{L}_W & = & - \frac{g}{\sqrt{2}} \overline{Q} \gamma^\mu (V^L_{Qq}P_L + V^R_{Qq}P_R )q W^+_\mu \nonumber\\
&& - \frac{g}{\sqrt{2}} \overline{q} \gamma^\mu (V^L_{qQ}P_L + V^R_{qQ}P_R )Q W^+_\mu   + H.c.\nonumber \\
\mathcal{L}_Z&=& - \frac{g}{2 c_W} \overline{q} \gamma^\mu (\pm X^L_{qQ}P_L \pm X^R_{qQ}P_R )QZ_\mu + H.c  \nonumber \\
\mathcal{L}_{h^0} & = & - \frac{g m_T}{2 M_W} \overline{t} ( Y^L_{htT} P_L + Y^R_{htT}P_R)Th^0 \nonumber \\
&&- \frac{gm_B}{2M_W} \overline{b} (Y^L_{hbB} P_L+ Y^R_{hbB} P_R ) B h^0 + H.c.\nonumber\\
\mathcal{L}_{H^0} & = &   -\frac{g m_T}{2 M_W}  \overline{t} ( Y^L_{HtT} P_L + Y^R_{HtT}P_R)T H^0 \nonumber \\
&&- \frac{gm_B}{2M_W}   \overline{b} (Y^L_{HbB} P_L+ Y^R_{HbB} P_R ) B h^0 + H.c. \nonumber\\
\mathcal{L}_{A} & = &   i\frac{g m_T}{2 M_W}  \overline{t} ( Y^L_{AtT} P_L - Y^R_{AtT}P_R)T A \nonumber \\
&&- i\frac{gm_B}{2M_W}  \overline{b} (Y^L_{AbB} P_L -  Y^R_{AbB} P_R ) B A + H.c. \nonumber \\
\mathcal{L}_{H^+} &=& - \frac{g m_T}{\sqrt{2}M_W}\overline{T} (\cot\beta Z^L_{Tb} P_L + \tan\beta Z^R_{Tb} P_R ) b H^+\nonumber\\
&&  - \frac{g m_B}{\sqrt{2}M_W}\overline{t} (\cot\beta Z^L_{tB} P_L + \tan\beta Z^R_{tB} P_R ) B H^+ +H.c,
\end{eqnarray}

\begin{eqnarray}
\begin{array}{c|cccc}
&	V_{tB}^L &V_{tB}^R  &	V_{bY}^L &V_{bY}^R  \\ \hline 
(B)	  & s_L  e^{i\phi}  &0 &-&-\\ 
(TB) & c_L^u s_L^d e^{i\phi_d}- s_L^u c_L^d e^{i\phi_u}  &-s_R^u c_R^d e^{i\phi_u}&-&-\\
(BY)  & s_L  e^{i\phi}  &0& -s_L e^{i\phi} &-s_R e^{i\phi}  \\
(XTB)  & (c_L^u s_L^d- \sqrt{2}s_L^u c_L^d) e^{i\phi} & -\sqrt{2}s_R^u c_R^d e^{i\phi} &- &-  \\
(TBY)  & (c_L^u s_L^d- \sqrt{2}s_L^u c_L^d) e^{i\phi} & -\sqrt{2}s_R^u c_R^d e^{i\phi} & -\sqrt{2}s_L^d  e^{i\phi} & -\sqrt{2}s_R^d  e^{i\phi}
\end{array} \nonumber
\end{eqnarray}\captionof{table}{Light-heavy couplings to the $W$ boson.}	

\begin{eqnarray}
\begin{array}{c|cc}
& 		X_{bB}^L &X_{bB}^R  \\ \hline

(B)	& c_L s_L e^{i\phi} &0  \\
(TB)	&0 &-s_R^d c_R^d e^{i\phi_d} \\
(BY)	&  2 c_L s_L e^{i\phi} &  c_R s_R e^{i\phi}\\
(XTB)	& -s_L^d c_L^d e^{i\phi} &  -2s_R^d c_R^d e^{i\phi}\\
(TBY)	&  s_L^d c_L^d e^{i\phi} & 0
\end{array} \nonumber
\end{eqnarray}\captionof{table}{Light-heavy couplings to the $Z$ boson.}

\begin{eqnarray}
\begin{array}{c|ccc }
&  	Y_{hbB}^L & 	Y_{HbB}^L& 	Y_{AbB}^L    \\ \hline
(B)	& (s_{\beta\alpha}  -   c_{\beta\alpha} \tan\beta)\frac{m_b}{m_B} c_L s_L e^{i\phi} &  (c_{\beta\alpha}  +  s_{\beta\alpha} \tan\beta)\frac{m_b}{m_B} c_L s_L e^{i\phi}   & \tan\beta\frac{m_b}{m_B}c_L s_L e^{i\phi}   \\ 
(TB) & (s_{\beta\alpha}  -   c_{\beta\alpha} \tan\beta) s_R^d c_R^d e^{i\phi_d} &  (c_{\beta\alpha}  +  s_{\beta\alpha} \tan\beta) s_R^d c_R^d e^{i\phi_d}  & \tan\beta s_R^d c_R^d e^{i\phi_d}   \\ 
(BY)	& (s_{\beta\alpha}  -    c_{\beta\alpha} \tan\beta) c_R s_R e^{i\phi} &  (c_{\beta\alpha}  +  s_{\beta\alpha} \tan\beta)c_R s_R e^{i\phi}   & \tan\beta c_R s_R e^{i\phi}\\
(XTB)	& (s_{\beta\alpha}  -   c_{\beta\alpha} \tan\beta)\frac{m_b}{m_B} s_L^d c_L^d e^{i\phi} &  (c_{\beta\alpha}  +  s_{\beta\alpha} \tan\beta)\frac{m_b}{m_B} s_L^d c_L^d e^{i\phi}   & \tan\beta\frac{m_b}{m_B} s_L^d c_L^d e^{i\phi}    \\
(TBY)	& (s_{\beta\alpha}  -   c_{\beta\alpha} \tan\beta)\frac{m_b}{m_B} s_L^d c_L^d e^{i\phi} &  (c_{\beta\alpha}  +  s_{\beta\alpha} \tan\beta)\frac{m_b}{m_B} s_L^d c_L^d e^{i\phi}   & \tan\beta\frac{m_b}{m_B} s_L^d c_L^d e^{i\phi}    
\end{array} \nonumber
\end{eqnarray}\captionof{table}{Light-heavy left couplings of Bottom quarks to the triplets Higgs \{$h,H,A$\}.}	

\begin{eqnarray}
\begin{array}{c|ccc }
&  	Y_{hbB}^R & 	Y_{HbB}^R & 	Y_{AbB}^R   \\ \hline
(B)	& (s_{\beta\alpha}  -    c_{\beta\alpha} \tan\beta) c_L s_L e^{i\phi} &  (c_{\beta\alpha}  +  s_{\beta\alpha} \tan\beta) c_L s_L e^{i\phi}   & \tan\beta c_L s_L e^{i\phi}   \\ 
(TB) & (s_{\beta\alpha}  -    c_{\beta\alpha} \tan\beta) \frac{m_b}{m_B} s_R^d c_R^d e^{i\phi_d} &  (c_{\beta\alpha}  +  s_{\beta\alpha} \tan\beta) \frac{m_b}{m_B} s_R^d c_R^d e^{i\phi_d}  &\tan\beta \frac{m_b}{m_B} s_R^d c_R^d e^{i\phi_d}\\ 
(BY)	& (s_{\beta\alpha}  -   c_{\beta\alpha} \tan\beta)\frac{m_b}{m_B} c_R s_R e^{i\phi} &  (c_{\beta\alpha}  +  s_{\beta\alpha} \tan\beta)\frac{m_b}{m_B} c_R s_R e^{i\phi}   & \tan\beta\frac{m_b}{m_B} c_R s_R e^{i\phi}\\
(XTB)	& (s_{\beta\alpha}  -   c_{\beta\alpha} \tan\beta)  s_L^d c_L^d e^{i\phi} &  (c_{\beta\alpha}  +  s_{\beta\alpha} \tan\beta)  s_L^d c_L^d e^{i\phi}   & \tan\beta  s_L^d c_L^d e^{i\phi}    \\
(TBY)	& (s_{\beta\alpha}  -   c_{\beta\alpha} \tan\beta)  s_L^d c_L^d e^{i\phi} &  (c_{\beta\alpha}  +  s_{\beta\alpha} \tan\beta)  s_L^d c_L^d e^{i\phi}   & \tan\beta  s_L^d c_L^d e^{i\phi}    
\end{array} \nonumber
\end{eqnarray}\captionof{table}{Light-heavy right couplings of Bottom quarks to the triplets Higgs \{$h,H,A$\}.}

\begin{eqnarray}
\begin{array}{c|ccccc}
& Z^L_{bY} & Z^R_{bY}& Z^L_{tB} && Z^R_{tB} \\ \hline 
(B)& -  &-& s_L && 0  \\ 
(BY)&  s_R e^{-i\phi}   &0&  0 &&0   \\ 
(TB)&- &-&  \frac{m_t}{ m_B}      \left[c_L^u s_L^d e^{ i\phi_d}  +    (  s_R^u{}^2    -  s_L^u{}^2)  \frac{c_L^d}{ s_L^u} e^{i\phi_u} \right] && c_L^u s_L^d e^{ i\phi_d}  +  (   s_L^d{}^2 - s_R^d{}^2)  \frac{ s_L^u}{c_L^d} e^{i\phi_u}    \\  
(TBY)&   c_L^u & 0 & s_L^d &&0
\end{array} 
\nonumber
\end{eqnarray}	
\captionof{table}{Light-heavy couplings to the charged Higgs.}
	
	
	\clearpage
	\bibliographystyle{JHEP}
	\bibliography{main}

\end{document}